\documentclass[iop]{emulateapj} 
\usepackage{graphicx}
\usepackage{apjfonts}
\usepackage{color}

\newcommand{\be}{\begin{equation}}
\newcommand{\ee}{\end{equation}}
\newcommand{\ba}{\begin{eqnarray}}
\newcommand{\ea}{\end{eqnarray}}

\newcommand{\mic}{\mu{\rm m}}

\newcommand{\kms}{\rm km\,s^{-1}}

\newcommand{\Mpc}{\rm Mpc}

\def\ls{\mathrel{\hbox{\rlap{\hbox{\lower4pt\hbox{$\sim$}}}\hbox{$<$}}}}
\def\gs{\mathrel{\hbox{\rlap{\hbox{\lower4pt\hbox{$\sim$}}}\hbox{$>$}}}}
\shorttitle{The SF--radius relation, slow quenching and pre-processing}
\shortauthors{Haines et al.}
\begin{document}
\title{LoCuSS: The slow quenching of star formation in cluster galaxies and the need for pre-processing}

\author{
  C.\,P.\ Haines\altaffilmark{1,2,3},
  M.\,J.\ Pereira\altaffilmark{2},
  G.\,P.\ Smith\altaffilmark{3},
  E.\ Egami\altaffilmark{2},
  A.\ Babul\altaffilmark{4},
  A.\ Finoguenov\altaffilmark{5,6},
  F.\ Ziparo\altaffilmark{3},
  S.\,L.\ McGee\altaffilmark{3,7},\\
  T.\,D.\ Rawle\altaffilmark{2,8},
  N.\ Okabe\altaffilmark{9}, 
  S.\,M. Moran\altaffilmark{10} 
\\
}

\altaffiltext{1}{Departamento de Astronom\'{i}a, Universidad de Chile, Casilla 36-D, Correo Central, Santiago, Chile; cphaines@das.uchile.cl} 
\altaffiltext{2}{Steward Observatory, University of Arizona, 933 North Cherry Avenue, Tucson, AZ 85721, USA} 
\altaffiltext{3}{School of Physics and Astronomy, University of Birmingham, Edgbaston, Birmingham, B15 2TT, UK}
\altaffiltext{4}{Department of Physics and Astronomy, University of Victoria, 3800 Finnerty Road, Victoria, BC, V8P 1A1, Canada}
\altaffiltext{5}{Department of Physics, University of Helsinki, Gustaf H\"{a}llstr\"{o}min katu 2a, FI-0014 Helsinki, Finland}
\altaffiltext{6}{Center for Space Science Technology, University of Maryland Baltimore County, 1000 Hilltop Circle, Baltimore, MD 21250, USA}
\altaffiltext{7}{Leiden Observatory, Leiden University, PO Box 9513, NL-2300 RA Leiden, the Netherlands}
\altaffiltext{8}{European Space Astronomy Centre, ESA, Villanueva de la Ca\~{n}ada, E-28691 Madrid, Spain}
\altaffiltext{9}{Department of Physical Science, Hiroshima University, 1-3-1 Kagamiyama, Higashi-Hiroshima, Hiroshima 739-8526, Japan}
\altaffiltext{10}{Smithsonian Astrophysical Observatory, 60 Garden Street, Cambridge, MA 02138, USA}

\
\begin{abstract}
We present a study of the spatial distribution and kinematics of star-forming galaxies in 30 massive clusters at $0.15{<}z{<}0.30$,  
combining wide-field {\em Spitzer} 24$\mu$m and {\em GALEX} NUV imaging with highly-complete spectroscopy of cluster members. 
The fraction ($f_{SF}$) of star-forming cluster galaxies rises steadily with cluster-centric radius, increasing fivefold by $2\,r_{200}$, but remains well below field values even at $3\,r_{200}$. This suppression of star formation at large radii cannot be reproduced by models in which star formation is quenched in infalling field galaxies only once they pass within $r_{200}$ of the cluster, but is consistent with some of them being first pre-processed within galaxy groups. 
Despite the increasing $f_{SF}$--radius trend, the surface density of star-forming galaxies actually declines steadily with radius, falling ${\sim}15{\times}$ from the core to $2\,r_{200}$. This requires star-formation to survive within recently accreted spirals for 2--3\,Gyr to build up the apparent over-density of star-forming galaxies within clusters.  
The velocity dispersion profile of the star-forming galaxy population shows a sharp peak of 1.44\,$\sigma_{\nu}$ at $0.3\,r_{500}$, and is 10-35\% higher than that of the inactive cluster members at all cluster-centric radii, while their velocity distribution shows a flat, top-hat profile within $r_{500}$. All of these results are consistent with star-forming cluster galaxies being an infalling population, but one that must also survive ${\sim}0$.5--2\,Gyr beyond passing within $r_{200}$. 
By comparing the observed distribution of star-forming galaxies in the stacked caustic diagram with predictions from the Millennium simulation, we obtain a best-fit model in which SFRs decline exponentially on quenching time-scales of $1.73{\pm}0.25$\,Gyr upon accretion into the cluster.

\end{abstract}
\keywords{ galaxies: active --- galaxies: clusters: general ---
  galaxies: evolution --- galaxies: stellar content }

\section{Introduction}
\label{intro}

\setcounter{footnote}{10}

The ability of galaxies to continuously form stars depends strongly on their global environment, with isolated central galaxies primarily evolving as star-forming spirals, while ``red and dead'' early-type galaxies completely dominate the cores of rich clusters, producing the SF--density or SF--radius relations \citep[e.g.][]{kennicutt83,lewis}. Various physical mechanisms have been proposed over the years to remove (or consume) gas and quench star formation in spiral galaxies within massive clusters, such as ram-pressure or viscous stripping, starvation, harassment or tidal interactions \citep[for reviews see e.g.][]{boselli,haines07}. 

Clusters and their member galaxies however do not exist and evolve in isolation from the rest of the Universe. In $\Lambda$CDM models structure formation occurs hierarchically, meaning that as the most massive collapsed structures in the Universe, galaxy clusters 
 form latest and are also the most dynamically immature \citep[e.g.][]{boylankolchin,gao12}. Preferentially residing at the nodes of the complex filamentary web, they continually accrete dark matter (DM) halos hosting individual ${\sim}L^{*}$ galaxies (M$_{\rm DM}{\sim}10^{12}{\rm M}_{\odot}$) or galaxy groups (M$_{\rm DM}{\sim}10^{13-14}{\rm M}_{\odot}$). The most massive clusters have on average doubled in mass since $z{\sim}0.5$ \citep{boylankolchin}, while half of galaxies in local clusters have been accreted since $z{\sim}0.4$ \citep{berrier}.

To correctly interpret the observed evolutionary and radial trends in cluster galaxy properties, it is thus fundamental to place them in this cosmological context whereby star-forming galaxies are being continually accreted into the clusters and transformed. Moreover, it is also vital to consider projection effects as many spectroscopic cluster members are actually infalling galaxies physically located outside the virial radius, and this contribution varies strongly with projected cluster-centric radius ($r_{proj}$) and line-of-sight (LOS) velocity relative to the cluster redshift ($v_{los}{-}{\langle}v{\rangle}$).
This requires using cosmological simulations containing one or more massive clusters, and following the orbits and merger histories of the galaxies or sub-halos which are accreted into the cluster over time \citep[e.g.][]{mamon10}. This approach gained early support when \citet{balogh00}, \citet{diaferio01} and \citet{ellingson} were able to reproduce the observed radial population gradients of star-forming galaxies in clusters in terms of galaxies on their first infall into the cluster. 

The caustic diagram, which plots $v_{los}{-}{\langle}v{\rangle}$ versus $r_{proj}$, has been used to constrain the kinematics and accretion epochs of different cluster galaxy populations, as well as the constrain the masses, density profiles and dynamical states of the clusters themselves \citep[e.g.][]{moss,binggeli,biviano97,biviano13,biviano,gill,mahajan11,hernandez,muzzin,jaffe}.

This progress has permitted recent attempts to constrain the time-scales required to halt star formation in recently accreted cluster spirals, with results supporting gentle physical mechanisms (e.g. starvation) that slowly quench star-formation over a period of several Gyrs \citep{wolf,vonderlinden,delucia,wetzel13}, rather than more violent processes (e.g. mergers) that rapidly terminate star formation \citep[although see e.g.][]{balogh04,mcgee11,wijesinghe}. 

In \citet{haines09a} we estimated the composite radial population gradients in the fraction of star-forming galaxies ($f_{SF}$) in 22 massive clusters at $0.15{<}z{<}0.30$ from the Local Cluster Substructure Survey (LoCuSS\footnote{http://www.sr.bham.ac.uk/locuss/}) based on panoramic {\em Spitzer}/MIPS 24$\mu$m data. A steady systematic increase in $f_{SF}$ with cluster-centric radius was observed out to ${\sim}r_{200}$, similar to those found previously \citep{ellingson,lewis,weinmann}. By comparison to galaxies infalling and orbiting around massive clusters (M$_{200}{\ga}10^{15}\,{\rm M}_{\odot}$) from the Millennium Simulation \citep{springel_mill}, it was possible to approximately reproduce the radial population trends in the context of a simple infall model, in which star-forming field galaxies are accreted into the cluster and their star-formation rapidly quenched upon their first pericenter. The key limitation of this work was the lack of redshifts to identify cluster galaxies, such that we had to statistically account for the contamination for field galaxy interlopers when estimating the $f_{SF}(r)$. 

We have since completed ACReS (Arizona Cluster Redshift Survey\footnote{http://herschel.as.arizona.edu/acres/acres.html}) which provides highly-complete spectroscopy of cluster members for all 30 clusters from  LoCuSS\footnote{http://herschel.as.arizona.edu/locuss/locuss.html} with wide-field {\em Spitzer}/MIPS data. With this data, \citet{haines13} found the specific-SFRs of massive ($\mathcal{M}{\ga}10^{10}\,{\rm M}_{\odot}$) star-forming cluster galaxies within $r_{200}$ to be systematically 28\% lower than their counterparts in the field at fixed stellar mass and redshift, a difference significant at the 8.7$\sigma$ level. This is the unambiguous signature of star formation in most (and possibly all) massive star-forming galaxies being {\em slowly} quenched upon accretion into clusters, and was best fit by models in which their star formation rates decline exponentially on quenching time-scales in the range 0.7--2.0\,Gyr.

In this article we analyse the spatial distribution and kinematics of star-forming galaxies within the same set of 30 clusters, and by comparing with predictions from cosmological simulations, draw further independent constraints on the quenching time-scale.
In particular, we determine the radial surface density profile, $\Sigma(r)$, of star-forming cluster galaxies and show that it declines steadily with radius, falling ${\sim}15{\times}$ from the core to $2\,r_{200}$. We show that this simple observation provides powerful constraints for how long massive star-forming galaxies are able to continue forming stars once they are accreted into rich clusters, quickly ruling out models in which star-formation is rapidly halted in infalling spirals when they pass within $r_{200}$. We also re-examine the radial population gradients of star-forming galaxies ($f_{SF}$--radius relation) out to $3\,r_{200}$, where we find a shortfall of star-forming galaxies in comparison to the coeval field population that cannot be easily explained by purely cluster-related quenching mechanisms, indicating a need for galaxies being first pre-processed within infalling galaxy groups.

In \S\ref{sec:data} we present our observational data, and in \S\ref{sec:results} the main results. In \S\ref{simulation} we follow the infall and orbits of galaxies in the vicinity of massive galaxy clusters from the Millennium simulation, to predict their spatial distributions and kinematics as a function of accretion epoch. These model predictions are then compared to observations in \S\ref{sec:quenching}. We discuss the resultant constraints on the time-scales required to quench star formation in recently accreted galaxies and the need for pre-processing in 
\S\ref{sec:discuss} and summarize in \S\ref{sec:summary}. Throughout we assume \mbox{$\Omega_M{=}0.3$},
\mbox{$\Omega_\Lambda{=}0.7$} and \mbox{${\rm
    H}_0{=}70\,\kms\Mpc^{-1}$}.

\section{Data}
\label{sec:data}

LoCuSS is a multi-wavelength survey of X-ray luminous galaxy clusters at $0.15{\le}z{\le}0.3$ 
\citep{smith10a} drawn from the ROSAT All Sky Survey cluster catalogs 
\citep{bohringer}.  
The first 30 clusters from our survey
benefit from a particularly rich dataset, including:
Subaru/Suprime-Cam optical imaging \citep{okabe10}, {\em Spitzer}/MIPS
$24\mic$ maps, {\em Herschel}/PACS+SPIRE 100--500$\mu$m maps, 
{\em Chandra} and/or {\em XMM} X-ray data, 
GALEX UV data, and
near-infrared (NIR) imaging. 
All of these data embrace at least $25^{\prime}{\times}25^{\prime}$ fields-of-view centered on each cluster, 
and thus probe them out to 1--2 virial radii \citep{haines10,pereira10,smith10b}.  
These 30 clusters were selected from the parent sample simply on the 
basis of being observable by Subaru on the nights allocated to us 
\citep{okabe10}, and should therefore not suffer any gross biases towards 
(for example) cool core clusters, merging clusters etc. Indeed, \citet{okabe10} show that the sample is statistically indistinguishable from a volume-limited sample.

\subsection{Chandra/XMM X-ray imaging}

All but two (Abell 291, Abell 2345) of the 30 clusters have available deep {\em Chandra} data ($t_{exp}{=}$9--120\,ksec). 
Deprojected dark matter densities, gas densities and gas temperature profiles for each cluster were derived by fitting the phenomenological cluster models of \citet{ascasibar} to a series of annular spectra extracted for each cluster \citep{sanderson10}. The best-fitting cluster models were then used to estimate $r_{500}$, the radius enclosing a mean overdensity of 500 with respect to the critical density of the Universe at the cluster redshift \citep{sanderson09}. 

The $r_{500}$ value for Abell 689 is taken from \citet{giles}, as they separated the extended cluster X-ray emission from the central BL Lac, and for the clusters lacking {\em Chandra} data, the $r_{500}$ values are taken from the {\em XMM} analysis of \citet{martino}. For the 23 clusters in common with the joint {\em Chandra--XMM} analysis of the LoCuSS high-$L_{X}$ cluster sample of \citet{martino}, there is good consistency of the cluster radii with $\langle r_{500,Haines}/r_{500,Martino}\rangle{=}1.003{\pm}0.034$.

The {\em Chandra} data were also used to identify X-ray AGN as described in \citet{haines12}. The survey limit of six broad (0.3--7\,keV) X-ray counts results in on-axis sensitivity limits of $L_{X}{\le}1.0{\times}10^{42}\,{\rm erg\,s}^{-1}$ for X-ray AGN at the cluster redshift for all 28 systems \citep[Table 1 from][]{haines12}.

Deep {\em XMM} data was available for 23 systems, allowing other groups and clusters in the region to be identified. Each 0.5--2\,keV image is decomposed into unresolved and extended emission, following the wavelet technique of \citet{finoguenov}. For each extended source, we attempt to identify the redshift of its associated group/cluster by examining the Subaru optical images for likely BCGs near the center of the X-ray emission and/or groups of galaxies with similar redshifts from ACReS within the X-ray contours. 

\subsection{Mid-infrared Observations}
\label{sec:coverage}

All 30 clusters were observed at $24\mic$ with MIPS \citep{rieke04} on board the {\em Spitzer Space
  Telescope}\footnote{This work is based in part on observations made
  with the Spitzer Space Telescope, which is operated by the Jet
  Propulsion Laboratory, California Institute of Technology under a
  contract with NASA (contract 1407).} (PID: 40872; PI: G.P. Smith).
The resulting 24$\mu$m mosaics were analysed with SExtractor \citep{bertin} producing catalogs which are on average 90\% complete to 400$\mu$Jy.

\begin{figure}
\centerline{\includegraphics[width=65mm]{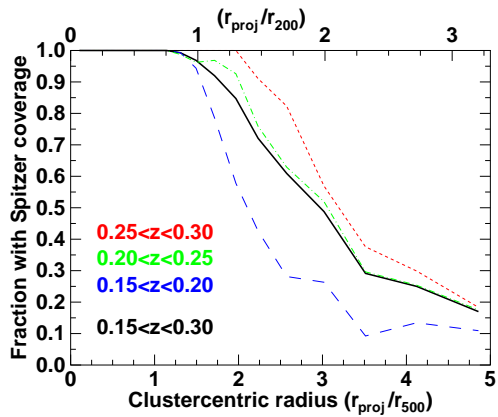}}
\caption{Radial coverage of the {\em Spitzer}/MIPS 24$\mu$m data.  Each curve shows the fraction of cluster members covered by our 24$\mu$m images ($25^{\prime}{\times}25^{\prime}$ field-of-view)  as a function of cluster-centric radius in units of $r_{500}$, for clusters in the following redshift ranges: $0.15{<}z{<}0.20$ ({\em long-dashed blue curves});  $0.20{<}z{<}0.25$ ({\em dot-dashed green curves});  $0.25{<}z{<}0.30$ ({\em short-dashed red curves});  $0.15{<}z{<}0.30$ ({\em solid black curves}).}
\label{coverage}
\end{figure}

Each cluster was observed across a fixed $25^{\prime}{\times}25^{\prime}$ field-of-view, resulting in the clusters being covered out to different cluster-centric radii in units of $r_{500}$, depending on their redshift and $r_{500}$ radius, as well as the orientation of the {\em Spitzer} images. 
Figure~\ref{coverage} shows that we probe out to larger cluster-centric radii for the highest redshift clusters ($0.25{<}z{<}0.30$) than those in our lowest redshift bin ($0.15{<}z{<}0.20$). 
Averaging over the full redshift range (0.15--0.30; {\em solid black curve}), our 24$\mu$m coverage is essentially complete out to $r_{200}$, falling to ${\sim}4$5\% at $2\,r_{200}$, based on the conversion $r_{500} = 0.66\,r_{200}$ \citep{sanderson03}.  

\begin{table*}
\centering
\begin{tabular}{llr@{ }lr@{.}lcr@{}lcc} \hline
Cluster   & \multicolumn{1}{c}{${\langle}z{\rangle}$} &\multicolumn{2}{c}{$N_{z}$ (lit)}&\multicolumn{2}{c}{$r_{500}$}& $M_{500}$ & \multicolumn{2}{c}{$\sigma_{\nu}$} & $\!\!t_{exp}(NUV)\!\!$ & $\!\!m_{AB}(NUV)\!\!$ \\ 
Name           &       &  &        & \multicolumn{2}{c}{(Mpc)} &$\!\!\!(10^{14}\,{\rm M}_{\odot})\!\!\!\!$& \multicolumn{2}{c}{(km\,s$^{-1})\!$} & (sec) & (S/N=5) \\ \hline 
Abell 68      &  0.2510 & 194&(0)     & 0&955 & 3.193 & 1186&$_{-88}^{+89}$ &   6731 & 22.90 \\ 
Abell 115    &  0.1919 & 213&(36)   & 1&304 & 7.628 & 1219&$_{-71}^{+72}$ & 16217 & 23.36 \\ 
Abell 209    &  0.2092 & 393&(49)$^{c}$& 1&230 & 6.519 & 1369&$_{-67}^{+65}$ & 11616 & 23.76 \\ 
Abell 267    &  0.2289 & 230&(139)$^{a}$ & 0&994 & 3.515 & 1045&$_{-52}^{+52}$ & 11268 & 23.32 \\ 
Abell 291    &  0.1955 & 126&(0)& 0&868$^{e}\!$&2.259 & 704&$_{-82}^{+80}$ &   6059 & 23.16 \\ 
Abell 383    &  0.1887 & 266&(92)$^{d}$ & 1&049 & 3.958 &   950&$_{-63}^{+64}$ &   6117 & 23.16 \\ 
Abell 586    &  0.1707 & 247&(21)   & 1&150 & 5.117 &   933&$_{-54}^{+55}$ &   ---   &   ---  \\ 
Abell 611    &  0.2864 & 297&(7)     & 1&372 & 9.847 & 1039&$_{-67}^{+67}$ &   9928 & 23.01 \\ 
Abell 665    &  0.1827 & 359&(31)   & 1&381 & 8.975 & 1227&$_{-59}^{+59}$ &   9503 & 23.26 \\ 
Abell 689    &  0.2776 & 338&(153)$^{a}$ & 1&126$^{f}\!$&5.390&721&$_{-58}^{+57}$ & ---  &   ---  \\ 
Abell 697    &  0.2821 & 486&(141)$^{a}$ & 1&505 & 12.93 & 1268&$_{-58}^{+57}$ & 36858 & 23.89 \\ 
Abell 963    &  0.2043 & 466&(50)$^{a}$  & 1&275 & 7.226 & 1119&$_{-49}^{+49}$ & 29043 & 23.96 \\ 
Abell 1689  &  0.1851 & 857&(416)$^{a}$ & 1&501 & 11.55 & 1541&$_{-46}^{+46}$ &   7716 & 23.34 \\ 
Abell 1758  &  0.2775 & 471&(50)$^{a}$   & 1&376 & 9.835 & 1442&$_{-63}^{+64}$ & 20977 & 23.88 \\ 
Abell 1763  &  0.2323 & 423&(126)$^{a}$ & 1&220 & 6.522 & 1358&$_{-53}^{+52}$ & 13589 & 23.80 \\ 
Abell 1835  &  0.2524 &1083&(608)$^{a}$& 1&589 & 14.73 & 1485&$_{-35}^{+35}$ & 21827 & 23.50 \\ 
Abell 1914  &  0.1671 & 454&(65)$^{a}$   & 1&560 & 12.73 & 1055&$_{-44}^{+44}$ &   3804 & 23.17 \\ 
Abell 2218  &  0.1733 & 342&(49)   & 1&258 & 6.716 & 1245&$_{-42}^{+42}$ & 14617 & 23.73 \\ 
Abell 2219  &  0.2257 & 628&(297)$^{a}$ & 1&494 & 11.90 & 1332&$_{-58}^{+59}$ &   9886 & 23.42 \\ 
Abell 2345  &  0.1781 &405&(39)& 1&249$^{e}\!$&6.607&1000&$_{-43}^{+43}$&   9864 & 23.49 \\ 
Abell 2390  &  0.2291 & 517&(140) & 1&503 & 12.16 & 1372&$_{-62}^{+63}$ &   6819 & 22.35 \\ 
Abell 2485  &  0.2476 & 196&(0)     & 0&830 & 2.088 &   799&$_{-54}^{+54}$ &   ---   &   ---  \\ 
RXJ0142.0+2131&0.2771&204&(15)& 1&136 & 5.531 & 1123&$_{-63}^{+66}$ &  ---    &   ---  \\ 
RXJ1720.1+2638&0.1599&473&(114)$^{b}$&1&530& 11.92 &  938 &$_{-36}^{+37}$ &   3074 & 22.67 \\ 
RXJ2129.6+0005&0.2337&334&(78)$^{a}$& 1&227 & 6.648 &  879 &$_{-82}^{+82}$ & 36323 & 23.72 \\ 
ZwCl0104.4+0048 (Z348)&0.2526& 185&(1) &0&760$^{e}$&1.613&806&$_{-73}^{+74}$&14435&23.46 \\ 
ZwCl0823.2+0425 (Z1693)&0.2261& 337&(4) & 1&050 & 4.130 & 671&$_{-48}^{+50}$ & 14314 & 23.41 \\ 
ZwCl0839.9+2937 (Z1883)&0.1931& 173&(3) & 1&107 & 4.674 & 834&$_{-87}^{+89}$ & 21250 & 23.39 \\ 
ZwCl0857.9+2107 (Z2089)&0.2344& 147&(0) & 1&024 & 3.866 & 815&$_{-80}^{+79}$ & 20225 & 23.68 \\ 
ZwCl1454.8+2233 (Z7160)&0.2565& 157&(1) & 1&128 & 5.294 &  988&$_{-84}^{+84}$& 19850 & 23.58 \\ \hline 
\end{tabular}
\caption{The cluster sample. Column (1) Cluster name; col. (2) Mean redshift of cluster members; col. (3) Total number of spectroscopic cluster members (contribution taken from the literature, including new redshifts from \citet{rines13}$^{a}$, \citet{owers}$^{b}$, \citet{mercurio}$^{c}$ and \citet{geller}$^{d}$); col. (4) radius $r_{500}$ in Mpc. $^{e}$From \citet{martino}, $^{f}$From Giles et al. (2012); col. (5) Cluster mass $M_{500}$ in $10^{14}{\rm M}_{\odot}$ ; col. (6) Velocity dispersion of cluster members within $r_{200}$; col. (7) Total exposure time of GALEX NUV images; col. (8) NUV magnitude limit for ${\rm SNR}=5$} 
\label{clusterlist}
\end{table*}

\subsection{UV, optical and near-infrared data}

Wide-field $J$- and $K$-band near-infrared imaging was obtained for all 30 clusters using either WFCAM on the 3.8-m United Kingdom Infrared Telescope (UKIRT)\footnote{UKIRT is operated by the Joint Astronomy Centre on behalf of the Science and Technology Facilities Council of the United Kingdom.} ($52^{\prime}{\times}52^{\prime}$ field-of-view; 26/30 clusters) or NEWFIRM on the 4.0-m Mayall telescope at Kitt Peak\footnote{Kitt Peak National Observatory, National Optical Astronomy Observatory, which is operated by the Association of Universities for Research in Astronomy (AURA) under cooperative agreement with the National Science Foundation.} ($27^{\prime}{\times}27^{\prime}$; 4/30 clusters), in each case reaching depths of $K{\sim}19$, $J{\sim}21$.

 Wide-field deep UV imaging from the {\em Galaxy Evolution Explorer} (GALEX) satellite was obtained for 26/30 clusters, primarily through the Cycle 4 (GI4-090; PI. G.P. Smith) and Cycle 6 (GI6-046; PI. S. Moran) Guest Investigator Programs. The Cycle 4 program provided far-ultraviolet (FUV) and near-ultraviolet (NUV) imaging for 14 clusters ($t_{exp}{=}3$.2--13.6\,ksec), while comparable FUV+NUV data were obtained for 7 more clusters ($t_{exp}{=}3$.4--29.0\,ksec) from the {\em GALEX} science archive. Sixteen clusters were observed in Cycle 6, including 8 systems not previously observed, but this provided only deep NUV imaging ($t_{exp}{=}2$.9--36.3\,ksec), the operations of the FUV camera having previously been suspended. The GALEX instrument has a circular field of view of radius 0.55\,deg, ensuring full ultraviolet coverage for galaxies in our near-infrared WFCAM fields. 
The total NUV exposure times and 5$\sigma$ magnitude limits, after correcting for Galactic extinction as in \citet{wyder}, are shown in Table~\ref{clusterlist}. 

Optical photometry in the $ugriz$ bands were taken from the Sloan Digital Sky Survey, using the dereddened SDSS model magnitudes. Twenty-six of the 30 clusters lie within the DR-10 footprint, while 23 have both SDSS $ugriz$ and deep GALEX NUV photometry, allowing star-forming galaxies to be identified from their blue $NUV{-}r$ colors.

\subsection{MMT/Hectospec spectroscopy}
\label{spectroscopy}

We have recently completed ACReS (the Arizona Cluster Redshift Survey; Pereira et al. 2015 in preparation) a long-term spectroscopic programme to observe our sample of 30 galaxy clusters with MMT/Hectospec. Target galaxies are primarily $K$-band selected down to a limit of $m_{K}^{*}(z_{cl}){+}1.5$ or fainter (depending on the number of targets produced), to produce an approximately stellar mass-limited sample down to $\mathcal{M}{\sim}2{\times}10^{10}\,{\rm M}_{\odot}$. Higher priorities are given to target galaxies also detected at 24$\mu$m to obtain a virtually complete census of obscured star formation in the cluster population. Further details of the survey aims and targetting strategy are given in \citet{haines13}. Eleven of our 30 clusters were also observed by the Hectospec Cluster Survey \citep[HeCS;][]{rines13}, providing redshifts for an additional 971 cluster members. 
Redshifts for a further 112, 92 and 49 members of clusters RXJ1720.1+2638, Abell 383 and Abell 209 are included from \citet{owers}, \citet{geller} and \citet{mercurio} respectively.  
Table~\ref{clusterlist} lists the number of spectroscopic members for each cluster, with the contributions taken from other published surveys indicated in parentheses, giving us a grand total of 10\,950 cluster members with redshifts. 
Averaging over all 30 systems, we achieve spectroscopic completeness levels of 66\% for $M_{K}{<}{-}23.10$ ($M_{K}^{*}{+}1.5$) cluster galaxies across the full WFCAM/NEWFIRM fields, rising to 80\% for those galaxies with {\em Spitzer} coverage and 96\% for those detected at 24$\mu$m.

The likelihood that a given galaxy was targetted for spectroscopy depends strongly on both its location with respect to the cluster center as well as its photometric properties ($K$-band magnitude, $J{-}K$ color, 24$\mu$m flux), as detailed in \citet{haines13}. To account for this, each galaxy is weighted by the inverse probability of it having being observed spectroscopically, following the approach of \citet{norberg}. 

\subsection{Identification of cluster members and field galaxy samples}
\label{data:members}

Members of each cluster are identified from the redshift versus projected cluster-centric radius plot as lying within the ``trumpet''-shaped caustic profile expected for galaxies infalling and subsequently orbiting within a massive virialized structure \citep{dunner}. For most systems, there is a strong contrast in phase-space density from inside to outside these caustics \citep{rines06}, making their visual identification relatively simple. 
The central redshift ${\langle}{z}{\rangle}$ and velocity dispersion $\sigma_{\nu}$ of each cluster (Table~\ref{clusterlist}) are iteratively measured for member galaxies within $r_{200}$ \citep[estimated as in][]{finn08}, using the biweight scale estimator \citep[$S_{BI}$;][]{beers}, with uncertainties estimated using bootstrap resampling. 

 The field galaxy sample was taken from the same dataset as the primary cluster galaxy sample, but were located in narrow redshift ranges on either side of the cluster, for which our spectroscopic survey ACReS should still be complete to $M^{*}_{K}+1.5$ \citep[for full details see][]{haines13}. Overall, 1398 coeval ($0.15{<}z{<}0.30$) field galaxies with $M_{K}{<}M_{K}^{*}{+}1.5$ and 24$\mu$m coverage were identified within these narrow redshift slices either side of the clusters (699 in front, 699 behind), after excluding regions where other X-ray galaxy groups had been previously detected from our {\em XMM} data.

\subsection{Stellar masses and SFRs}
\label{sec:lir}

Rest-frame UV--optical colors and absolute magnitudes were determined using the k-corrections of \citet{chilingarian}.
Stellar masses ($\mathcal{M}$) were estimated from the $K$-band luminosities using the linear relation between $K$-band stellar mass-to-light ratio and rest-frame $g-i$ color from \citet{bell}, adjusted by -0.15\,dex to be valid for a \citet{kroupa} IMF. 
Where SDSS photometry was unavailable we classified the galaxy as being either star-forming or passive according to whether it was 24$\mu$m or NUV detected or not, and adopted appropriate M/L ratios. 

For each 24$\mic$-detected galaxy with known redshift, its intrinsic bolometric luminosity ($L_{TIR}$) and rest-frame 24$\mu$m luminosity is estimated by comparison of its 24$\mu$m flux to the luminosity-dependent template infrared spectral energy distributions of \citet{rieke}. The latter is then converted to an obscured SFR using the calibration of \citet{rieke}
\begin{equation}
{\rm SFR}_{IR}{\rm (M}_{\odot}{\rm yr}^{-1}) = 7.8{\times}10^{-10}\,L(24{\mu}{\rm m}, L_{\odot})
\end{equation}
which is valid for either a \citet{kroupa} or \citet{chabrier} IMF. 
Our {\em Spitzer} data should be sensitive to galaxies with ongoing obscured star formation occuring at rates down to 2.0\,M$_{\odot}\,{\rm yr}^{-1}$ in our most distant clusters ($z{\sim}0.28$).

Local quiescent early-type galaxies are known to emit in the mid-infrared at levels much higher than expected from photospheric emission alone \citep{clemens}. This excess at 10--30$\mu$m has been shown to be due to silicate emission from the dusty circumstellar envelopes of mass-losing evolved AGB stars \citep{bressan} rather than residual ongoing star formation. We may thus worry that some of our cluster galaxies may be mistakenly classed as star-forming due to 24$\mu$m emission coming from TP-AGB stars.

The spectral energy distributions (SEDs) of evolved stellar populations including emission from dusty circumstellar envelopes peak at 10--20$\mu$m, but then drop rapidly at longer wavelengths \citep{piovan}, and so galaxies whose 24$\mu$m emission is due to TP-AGB stars should not be detected in our {\em Herschel}/PACS data, unlike normal star-forming galaxies whose infrared SEDs peak at 70--170$\mu$m \citep[e.g.][]{dale}.
For the 11 nearest clusters ($0.15{<}z{<}0.20$) in our sample, $>$98\% of galaxies with SFR$_{IR}{>}2.0\,{\rm M}_{\odot}{\rm yr}^{-1}$ and {\em Herschel}/PACS coverage were also detected at 100$\mu$m, while $>$99\% show clear H$\alpha$ emission in our ACReS MMT/Hectospec spectra, indicating that their 24$\mu$m emission is indeed due to ongoing star formation.

In a comparable {\em Spitzer}/MIPS analysis of 814 galaxies in the Shapley supercluster at $z{=}0.048$, sensitive to much lower obscured star formation rates (SFR$_{IR}{\sim}0.05\,{\rm M}_{\odot}{\rm yr}^{-1}$) \citet{haines11b} did find a significant population of quiescent (based upon a lack of H$\alpha$ emission) cluster galaxies detected at 24$\mu$m, but none with 24$\mu$m luminosities that would convert to an obscured SFR above 0.5\,M${_\odot}{\rm yr}^{-1}$. They also obtained a tight correlation (0.22\,dex) between the 24$\mu$m and the 1.4\,GHz radio luminosities for star-forming cluster galaxies, down to SFR$_{IR}{\sim}1.0\,{\rm M}_{\odot}{\rm yr}^{-1}$, consistent with both the mid-infrared and radio emission being due to star formation.

\section{Results}
\label{sec:results}

\begin{figure}
\centerline{\includegraphics[height=80mm]{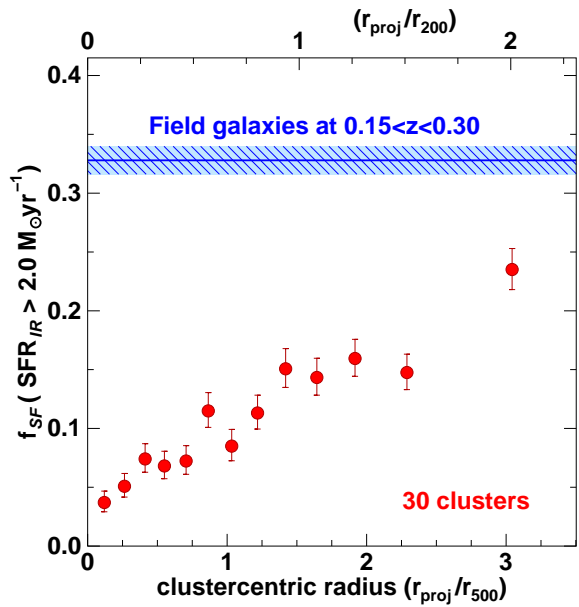}}
\caption{Radial population gradients for MIR-selected star forming galaxies from our stacked sample of 30 clusters. Red symbols show the fraction of massive ($\mathcal{M} > 2.0{\times}10^{10}\,{\rm M}_{\odot}$)  cluster galaxies with obscured star-formation at rates ${\rm SFR}_{IR} > 2.0\,{\rm M}_{\odot}\,{\rm yr}^{-1}$ as a function of projected cluster-centric radius ($r_{proj}/r_{500}$). The error bars indicate the uncertainties derived from binomial statistics calculated using the formulae of \citet{gehrels}. Each radial bin contains 400 cluster galaxies. The blue horizontal line indicates the corresponding fraction of field galaxies ($\mathcal{M}{>}2.0{\times}10^{10}\,{\rm M}_{\odot}$; $0.15{<}z{<}0.30$) with ${\rm SFR}_{IR} > 2.0\,{\rm M}_{\odot}\,{\rm yr}^{-1}$ and its $1\sigma$ confidence limits (shaded region). 
} 
\label{sf_radius}
\end{figure}
 
\subsection{Radial population gradients}

\begin{figure}
\centerline{\includegraphics[width=84mm]{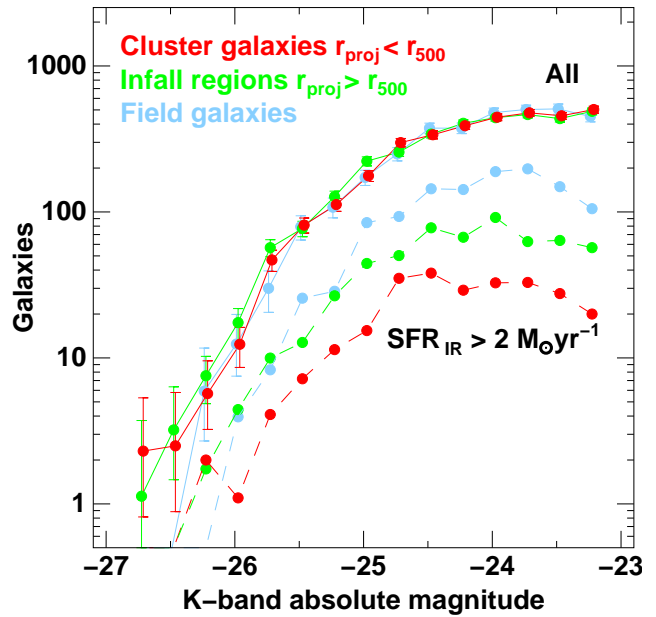}}
\caption{$K$-band luminosity functions (LFs) for cluster galaxies within $r_{500}$ ({\em red solid curve}), galaxies from the infall regions ($r_{proj}{>}r_{500}$; {\em green solid curve}) and the coeval field population ({\em light-blue solid curve}). The latter two are normalized to contain the same total number of $M_{K}{<}{-}23.1$ galaxies as the cluster galaxy population, and the points slightly shifted horizontally, to permit easier comparison of the LFs. Error bars indicate Poissonian uncertaintes based on \citet{gehrels}. The corresponding $K$-band LFs for star-forming galaxies with SFR$_{IR}{>}2.0\,{\rm M}_{\odot}{\rm yr}^{-1}$ are indicated by dashed curves.}
\label{lfs}
\end{figure}

Figure~\ref{sf_radius} shows the fraction of massive ($\mathcal{M}{>}2.0{\times}10^{10}\,{\rm M}_{\odot}$) cluster galaxies with obscured star formation occurring at rates \mbox{${\rm SFR}_{IR} > 2.0\,{\rm M}_{\odot}\,{\rm yr}^{-1}$} ($f_{SF}$) as a function of projected cluster-centric radius in units of $r_{500}$, across the full sample of 30 clusters. 
The fractions only consider cluster members covered by the {\em Spitzer} 24$\mu$m maps, and exclude X-ray AGN and QSOs, as their 24$\mu$m emission is usually dominated by dust heated by the active nucleus rather than star formation (e.g. Xu et al. 2015). BCGs are also excluded due to their unique evolutions \citep{lin}, and the direct link between BCG activity and the presence of cooling flows within clusters \citep{smith10a, rawle}. 

The fraction of obscured star-forming galaxies increases steadily with cluster-centric radius from $f_{SF}{\sim}0.04$ in the cluster core to $f_{SF}{\sim}0.23$ at $3.0\,r_{500}$ ($1.9\,r_{200}$). However even at these large radii the $f_{SF}$ remains well (${\sim}$1/3) below that seen in coeval field galaxies ($f_{SF}{=}0.33{\pm}0.01$; {\em blue line}). A simple linear extrapolation of the observed trend suggests that the $f_{SF}$ should reach that of the field galaxy population at ${\sim}4.5\,r_{500}$. However the limited extents of our {\em Spitzer} 24$\mu$m maps mean that we cannot establish whether this occurs or not. 

The fraction of star-forming cluster galaxies evolves very rapidly at these redshifts, with $f_{SF}{\propto}(1+z)^{7.6{\pm}1.1}$ \citep{haines13}. It is thus vital to ensure that this shortfall in star-forming cluster galaxies at large radii with respect to field values is not produced by a redshift bias between the two samples. This is certainly not the case here, as the mean redshifts of each radial bin for the cluster populations all lie in the range 0.217--0.241, while for the field population $\langle z\rangle=0.225$. 
There is a marginal redshift bias within our cluster sample, as the outer two radial bins have $\langle z\rangle = 0.241$, while the remaining bins all have mean redshifts in the range 0.217--0.231. This is due to our {\em Spitzer} data providing wider radial coverage (in terms of $r_{500}$) for the higher redshift systems, but it is likely only a marginal effect, artificially increasing the outer two $f_{SF}$ by $\la 1$0\% (or ${\sim}0$.02 in the figure). 

\begin{figure}
\centerline{\includegraphics[height=80mm]{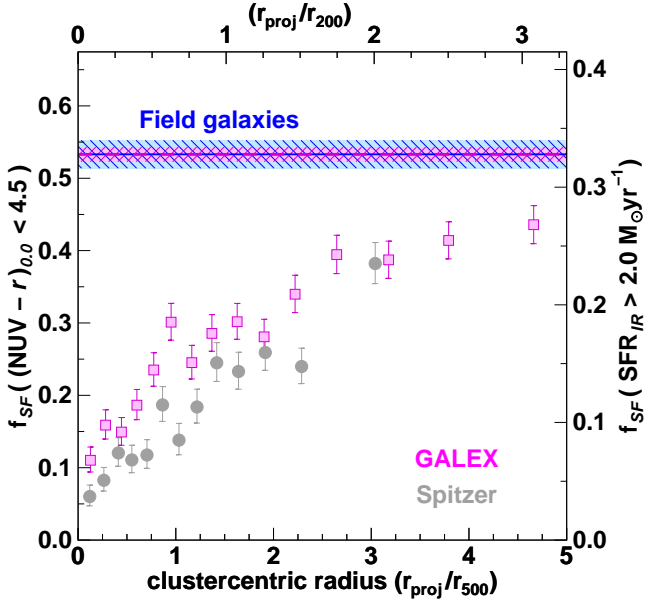}}
\caption{Comparison of radial population gradients for star-forming cluster galaxies selected in the ultraviolet and mid-infrared. Magenta squares show the fraction of massive ($\mathcal{M}{>}2.0{\times}10^{10}\,{\rm M}_{\odot}$) cluster galaxies with blue rest-frame UV--optical colors, $(NUV{-}r)_{0.0}{<}4.5$, indicating ongoing unobscured star-formation. Each radial bin contains 300 cluster galaxies. The magenta dashed line indicates the corresponding fraction of field galaxies ($\mathcal{M}{>}2.0{\times}10^{10}\,{\rm M}_{\odot}$; $0.15{<}z{<}0.30$) with $(NUV{-}r)_{0.0}{<}4.5$ and its $1\sigma$ confidence levels (magenta shaded region). Grey symbols show the corresponding fractions of cluster galaxies with ${\rm SFR}_{IR} > 2.0\,{\rm M}_{\odot}\,{\rm yr}^{-1}$ taken from Fig.~\ref{sf_radius}, scaled via the right-hand axis, such that the fraction of field galaxies with ${\rm SFR}_{IR} > 2.0\,{\rm M}_{\odot}\,{\rm yr}^{-1}$ (blue dashed line) coincides with the corresponding fraction selected in the NUV. 
}
\label{UV_sf_radius}
\end{figure}

A second possible explanation for the lower $f_{SF}$ among cluster galaxies would be if they were more massive on average than the field galaxy comparison sample, as $f_{SF}$ is known to decline with increasing stellar mass at fixed galaxy density \citep[e.g.][]{haines07}. 
However, the mean stellar masses of cluster galaxies for each radial bin in Fig.~\ref{sf_radius} are always within 0.04\,dex of that of the coeval field population. Moreover, the  $K$-band luminosity functions of cluster galaxies within $r_{500}$ (excluding BCGs), those in the infall regions ($r_{proj}{>}r_{500}$), and the field galaxy samples are all indistinguishable (Fig.~\ref{lfs}). We also note that due to the fixed SFR lower limit of 2.0\,M$_{\odot}\,{\rm yr}^{-1}$ used to define a star-forming galaxy, the $f_{SF}$ do not vary much with stellar mass ($K$-band luminosity). 
We therefore exclude secular quenching due to increased stellar masses among the cluster galaxy population as being responsible for their lower $f_{SF}$ (at all radii in Fig.~\ref{sf_radius}) with respect to the field. The progressive suppression in star-formation in moving from field to infall regions, and on to cluster environments is seen at all stellar masses (compare dashed curves in Fig.~\ref{lfs}).

The {\em GALEX} NUV data provides a complementary means of identifying star forming galaxies from their ultraviolet emission, and an opportunity to measure the SF--radius relation out to larger cluster-centric radii. Figure~\ref{UV_sf_radius} shows the composite radial population gradient ({\em magenta squares}) in the fraction of cluster galaxies having blue rest-frame UV--optical colors $(NUV{-}r)_{0.0}{<}4.5$, for the 23 clusters with deep {\em GALEX} NUV data and SDSS $ugriz$ photometry. The $(NUV{-}r)_{0.0}{=}4.5$ color limit lies in the middle of the UV--optical ``green valley'' \citep{wyder}, and allows passively-evolving galaxies to be efficiently excluded without losing dusty star-forming galaxies due to reddening \citep{haines08}. Optically-quiescent early-type galaxies with residual (or ``rejuvenated'') extended star formation in the form of rings or spiral arms should also still be recovered \citep{salim}. 

As before, the fraction of star-forming cluster galaxies increases steadily with cluster-centric radius from $f_{SF}{\sim}0.11$ in the cluster core to $f_{SF}{\sim}0.44$ at 4--5\,$r_{500}$ (${\sim}3\,r_{200}$). The fraction of star-forming cluster galaxies remains significantly below that seen in coeval field galaxies ($f_{SF}{=}0.53{\pm}0.01$; {\em magenta dashed line}), even out at ${\sim}5\,r_{500}$, the trend appearing to flatten off rather than continue upwards to field values.  
Again the two samples are confirmed to be coeval: each radial bin for the cluster population has a mean redshift in the range 0.217--0.236, while for field galaxies $\langle z\rangle =0.226$. 

To allow comparison between the NUV-based and 24$\mu$m-based SF--radius relations, the latter ({\em gray points}) is replotted from Fig.~\ref{sf_radius} after adjusting its vertical scale ({\em right hand axis}) to ensure that fractions of IR-selected and UV-selected star-forming field galaxies coincide on the plot. 
While both relations show the same steadily increasing trends with radius, the NUV-based SF--radius relation consistently lies above the re-scaled IR-based relation.

Looking at the IR-based SF-radius relation, it is tempting to suggest that the fraction of star-forming galaxies even falls to zero at the cluster core. However, we confirm that this is not the case. Further splitting each radial bin into four, the $f_{SF,IR}$ never fall below a floor value of ${\sim}$3--5\% in the cluster core. Similarly, the NUV-based SF-radius relation never falls below ${\sim}$8--15\% in the cluster core, when the radial bins are further sub-divided. 
A residual population of cluster galaxies with ongoing star formation exists at all radii. 
  
\begin{figure}
\centerline{\includegraphics[width=70mm]{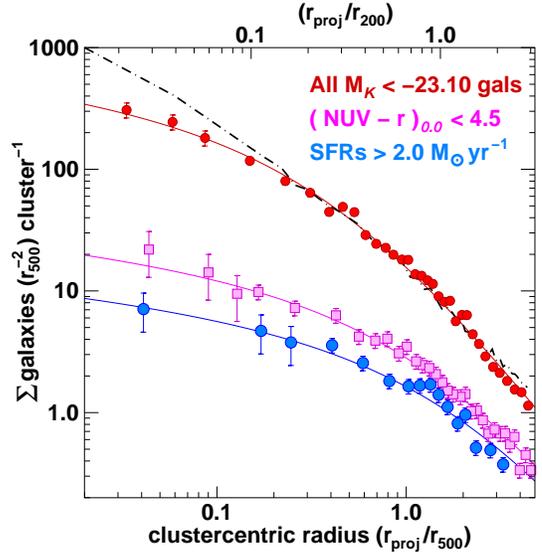}}
\caption{Composite galaxy surface density profile for {\em all} $M_{K}{<}M_{K}^{*}{+}1.5$ 
galaxies (BCGs excluded) from our sample of 30 $0.15{<}z{<}0.30$ clusters, as a function of projected cluster-centric radius ($r_{proj}/r_{500}$). Each symbol consists of 200 spectroscopic cluster members. The blue symbols show the corresponding observed surface density $\Sigma(r)$ of $M_{K}{<}M_{K}^{*}{+}1.5$  
galaxies with active obscured star-formation at rates SFR$_{IR}{>}2.0\,{\rm M}_{\odot}{\rm yr}^{-1}$ from the same ensemble cluster. The magenta squares show the radial profile for unobscured star-forming galaxies with $(NUV-r)_{0.0}{<}4.5$ from the 22 clusters that have SDSS $ugriz$ and deep GALEX NUV photometry. The error bars assume Poisson statistics. The corresponding best-fit NFW profiles are shown by the red, blue and magenta curves. 
The dot-dashed black curve shows the predicted surface density profiles for all ``spectroscopic'' member galaxies of the 75 most massive clusters from the Millennium simulation, scaled to fit our observed ensemble surface density profile of all galaxies. 
}
\label{log_profile}
\end{figure}
 
\subsection{Radial surface density profiles}

The spatial distribution of galaxies within clusters provide key constraints on the primary epoch at which they were accreted into the system, as well as the effects of continual cluster mass growth on their orbital parameters and tidal and ram-pressure stripping on their stellar masses. Figure~\ref{log_profile} shows the radial distribution of all $M_{K}<M_{K}^{*}{+}1.5$ 
galaxies in the stacked cluster out to ${\sim}4\,r_{500}$, excluding BCGs ({\em red points}). 

Large numerical DM simulations have found that DM halos are well described by a ``universal'' 2-parameter \citep[NFW;][]{nfw} density profile from scales of 10\,kpc out to 10\,Mpc \citep{frenk,gao12}. 
The NFW profile is characterized by a scale radius $r_{s}{=}r_{200}/c$, where $c$ is the concentration parameter. The three-dimensional density profile is given by $\rho(x)\propto x^{-1}(1+x)^{-2}$, where $x{=}r/r_{s}$, and $d\log\rho/d\log r = {-}2$ at $r{=}r_{s}$. 
The NFW model has been shown to provide excellent fits to stacked tangential shear profiles of massive clusters \citep{okabe13}, and the distribution of cluster galaxies \citep{lin04}. 
The surface density profile, $\Sigma(r)$, of cluster galaxies in our ensemble cluster sample can be well described by a projected NFW profile with a concentration parameter $c_{g}{=}3.01{\pm}0.16$ ({\em red curve}), consistent with the $c_{g}{=}2.90{\pm}0.22$ value obtained by \citet{lin04} or $c_{g}{=}2.7{\pm}0.7$ obtained by \citet{budzynski}, while \citet{muzzin07} obtained a higher concentration of $c_{g}{=}4.13{\pm}0.57$ for the stacked $K$-band number density profile of galaxies from 15 clusters at $0.2<z<0.55$. 
The $c_{g}$ value of 3.01 obtained here is still significantly lower than the $c_{WL,200}{=}4.22_{-0.36}^{+0.40}$ value for the concentration of the overall mass distribution obtained by \citet{okabe13} from their weak lensing analysis of 50 $z{\sim}0.2$ clusters, including many of the systems in our sample. 

\begin{figure}
\centerline{\includegraphics[width=84mm]{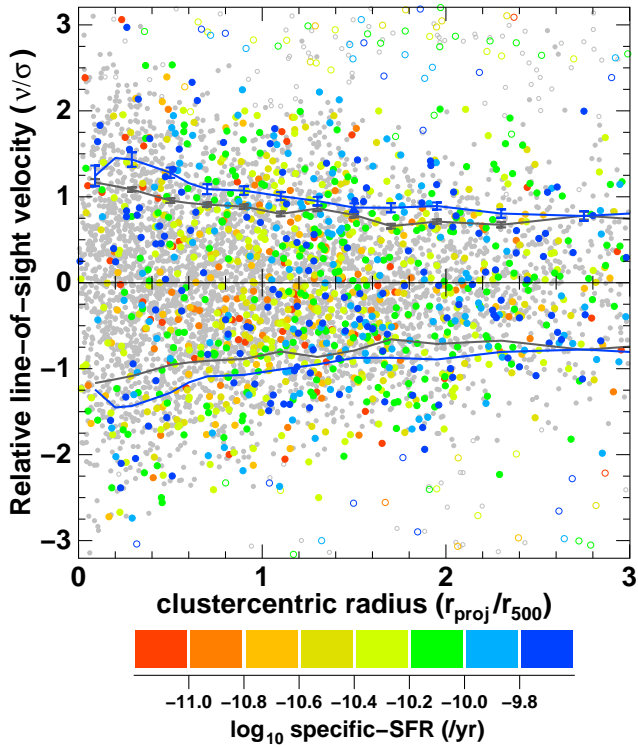}}
\caption{Stacked observed phase-space diagram,  $(\nu_{los}{-}\langle\nu\rangle)/\sigma_{\nu}$ versus $r_{proj}/r_{500}$, 
of galaxies combining all 30 clusters in our sample. Each gray solid dot represents a spectroscopic cluster member, while open points indicate non-cluster galaxies. Only those galaxies covered by our {\em Spitzer} data are plotted. 
Star-forming galaxies detected at 24$\mu$m are indicated by larger symbols, colored according to their specific-SFRs from red (${\rm sSFR}{<}10^{-11}$\,yr$^{-1}$) to blue (${\rm sSFR}{>}10^{-10}$\,yr$^{-1}$).  The blue curve and error bars indicates the 1-$\sigma$ velocity dispersion profile of the 24$\mu$m-detected cluster members. The uncertainty in each $\sigma(r)$ value is estimated by bootstrap resampling the galaxies in that radial bin. The black curve shows the corresponding radial profile for the remaining inactive cluster members.} 
\label{sf_zspace}
\end{figure}

The surface density, $\Sigma(r)$, of massive ($M_{K}{<}M_{K}^{*}{+}1.5$) cluster galaxies with obscured ${\rm SFR}_{IR} {>} 2.0\,{\rm M}_{\odot}{\rm yr}^{-1}$ ({\em blue points}) declines steadily with radius from ${\sim}7$\,galaxies\,$r_{500}^{-2}$\,cluster$^{-1}$ in the cluster cores to ${\sim}0$.4 by ${\sim}3\,r_{500}$. 
These $\Sigma(r)$ include corrections for spectroscopic incompleteness ({\S}~\ref{spectroscopy}) and the radial variation in coverage by the 24$\mu$m images ({\S}~\ref{sec:coverage}).  
Even though the {\em fraction} of star-forming galaxies is falling to close to zero in cluster cores, clusters mark over-densities in the spatial distribution of star-forming galaxies in the plane of the sky. This is simply due to the cuspy density profile of the global cluster galaxy population more than compensating for the steady decrease in $f_{SF}$ when approaching the cluster core.  
The surface density of $M_{K}{<}M_{K}^{*}{+}1.5$ cluster galaxies with \mbox{$(NUV-r)_{0.0}<4.5$} ({\em magenta squares}) shows a similar radial profile, but is marginally steeper at small radii ($r_{proj} \la 0.2\,r_{500}$).
We will show in {\S}~\ref{sec:profiles} that these steadily declining trends in $\Sigma(r)$ imply that star formation must survive within recently accreted spirals for several Gyr to build up the apparent over-densities of star-forming cluster galaxies. 

\subsection{Dynamical analysis of star-forming cluster galaxies}

\begin{figure}
\centerline{\includegraphics[width=84mm]{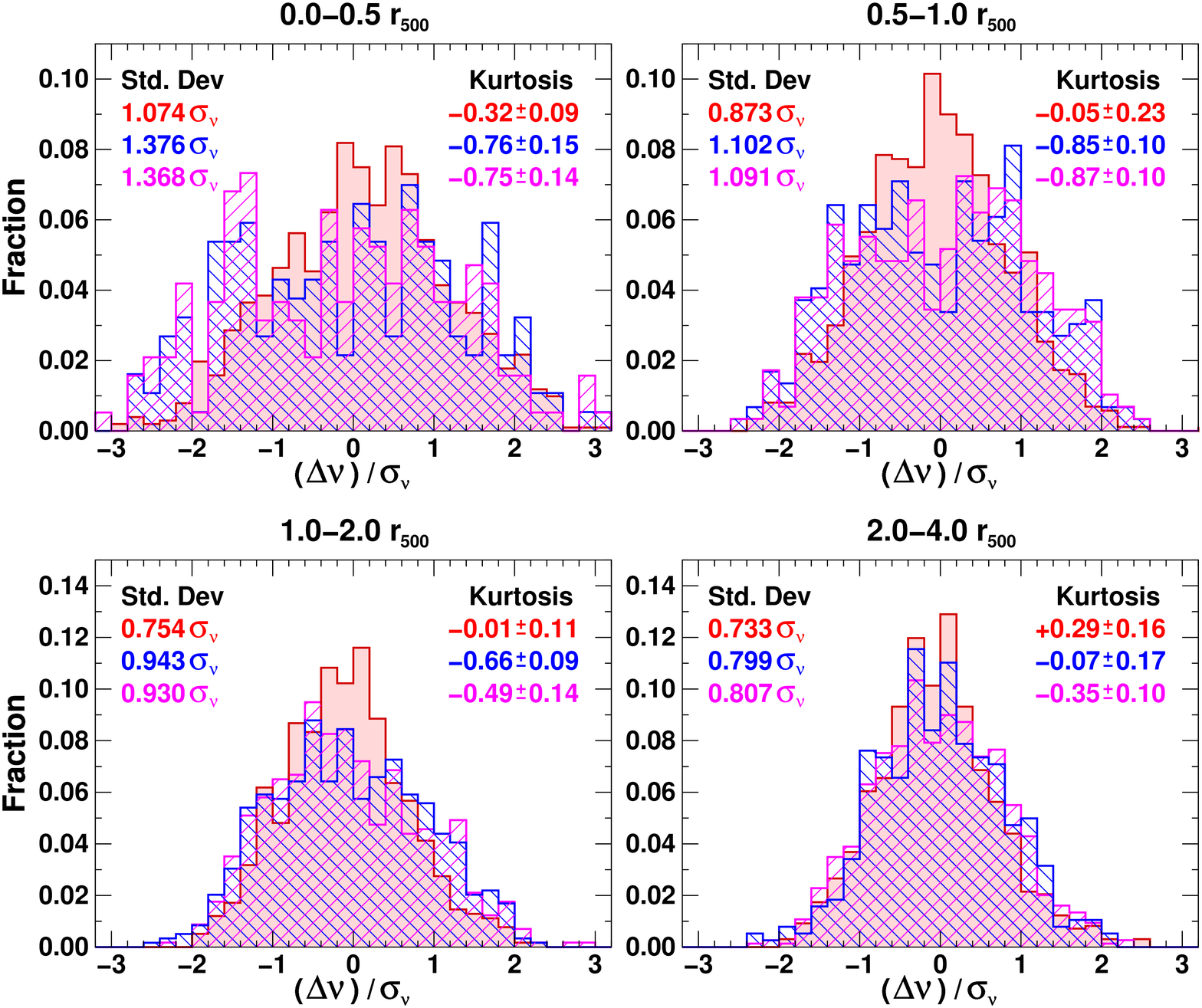}}
\caption{Stacked relative LOS velocity distributions of inactive galaxies ({\em red solid histograms}), 24$\mu$m-detected ({\em blue striped histogram}) and UV-selected ($NUV-r<4.5$; {\em magenta striped histogram}) star-forming galaxies from our ensemble cluster, in four bins of projected radius. The standard deviations (in units of $\sigma_{\nu}$) and kurtosis values of each distribution are indicated.} 
\label{vel_dists}
\end{figure}

Figure~\ref{sf_zspace} shows the stacked caustic diagram of all 30 clusters, in which the projected radius of each cluster member is normalized by the {\em Chandra}-based $r_{500}$ of that cluster, and the LOS velocities are scaled in units of $\sigma_{\nu}$. 
The 24$\mu$m-detected star-forming galaxies ({\em colored points}) do not have the same spatial distribution within the caustic diagram as the remaining inactive cluster galaxy population ({\em gray points}). They appear to preferentially lie along the caustics, indicative of an infalling population. They also show a concentration at $r_{proj}{\sim}1$.0--1.$5\,r_{500}$, covering the full velocity range within the caustics. The fall off in numbers towards larger radii is a selection effect due to the decline in {\em Spitzer} 24$\mu$m coverage beyond $2\,r_{500}$.
Star-forming galaxies do not entirely avoid the central region with low cluster-centric radii ($r_{proj}{<}0.4\,r_{500}$) and relative LOS velocities ($|\Delta \nu/\sigma_{\nu}|{\la}0.80$), as X-ray AGN appear to do \citep{haines12}, but their frequency certainly drops off here, in marked contrast to the inactive cluster galaxy population. Some of the star-forming galaxies in these central regions of phase-space will likely appear here due to projection effects, being located along the line-of-sight of the cluster but physically still well outside $r_{200}$ (see Fig.~\ref{quenching_caustics}).

The velocity dispersion of the star-forming cluster galaxy population ({\em blue curve}) is 10--35\% higher than that of the inactive cluster galaxies ({\em gray curve}) at all radii.  
Averaging over all galaxies within $r_{500}$ ($2\,r_{500}$), star-forming galaxies have absolute LOS velocity offsets which are 26.4\% (24.9\%) higher than their passive counterparts at the same cluster-centric radius, a result significant at the 8.0$\sigma$ (10.7$\sigma$) level.  

The same trends are obtained when selecting star-forming cluster galaxies according to their rest-frame $NUV-r$ color. 
Their velocity dispersion remains 10--35\% higher than that of inactive cluster galaxies out to 2.5--3\,$r_{500}$, although at larger radii they become indistinguishable. 

The relative LOS velocity distributions of star-forming (both 24$\mu$m-detected and UV-selected) and inactive cluster galaxies for the stacked LoCuSS cluster sample are shown in Figure~\ref{vel_dists}, in four bins of projected cluster-centric radius. The LOS velocity distributions of inactive galaxies (not detected at 24$\mu$m and having $(NUV-r)_{0.0}>4.5$) can be approximately described as a Gaussian at all radii. The distributions of star-forming galaxies in the two inner radial bins ($r_{proj}{<}r_{500}$) instead appear more consistent with a flat, top-hat profile than a Gaussian, including an relative excess of star-forming galaxies at $|\Delta \nu/\sigma_{\nu}|>1.2$ in comparison to the inactive population. At 0.5--1.0\,$r_{500}$ there is even marginal evidence a central dip in the LOS velocity distribution of star-forming galaxies. The kurtosis, $\gamma_{2}=[\frac{1}{N}\sum_{i=1}^{N}(\nu_{i}-\bar{\nu})^{4} / (\frac{1}{N}\sum_{i=1}^{N}(\nu_{i}-\bar{\nu})^{2})^{2}] -3$ of the LOS velocity distributions of star-forming galaxies ($\gamma_{2}{\sim}-0.81$) is significantly lower than that expected for a Gaussian distribution ($\gamma_{2}{=}0.0$) at ${>}5\sigma$ level in both inner radial bins, and closer to expectations for a uniform top-hat distribution ($\gamma_{2}{=}-1.2$). At large cluster-centric radii (2.0--4.0\,$r_{500}$), the velocity distribution of star-forming galaxies is consistent with a Gaussian function, and is almost indistinguishable from that of the inactive population, albeit with a marginally (${\sim}9$\%) higher dispersion.

\begin{figure}
\centerline{\includegraphics[width=84mm]{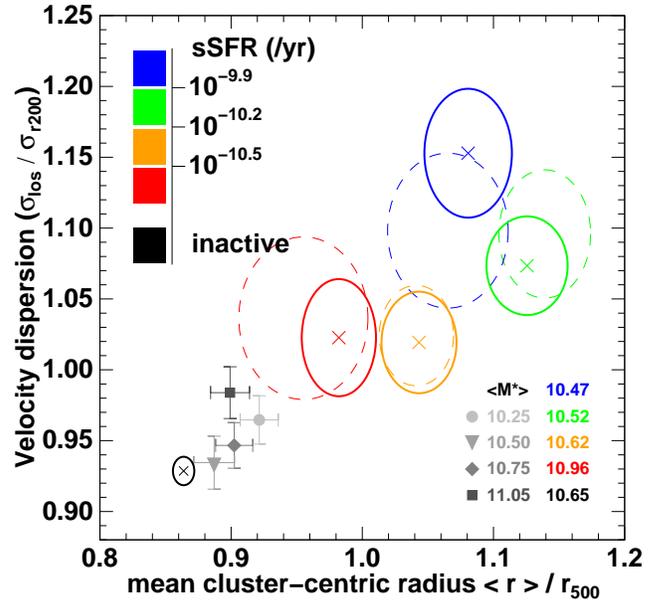}}
\caption{Variation in the spatial distribution and kinematics of cluster galaxies as a function of their specific-SFR and stellar mass. Colored crosses indicate the mean cluster-centric radii (${\langle}r_{proj}/r_{500}{\rangle}$) and LOS velocity dispersions ($\sigma_{los}(r_{proj}{<}2.0\,r_{500})/\sigma_{\nu}$) of star-forming cluster galaxies split into four bins of specific-SFR. Error ellipses show the 1$\sigma$ uncertainties in both values. The black cross indicates the corresponding values for inactive cluster galaxies not detected with {\em Spitzer}. Only cluster galaxies within $2\,r_{500}$ are included within each sub-sample. The mean stellar mass, $\langle \log (\mathcal{M}/{\rm M}_{\odot})\rangle$, of each sub-sample is indicated in the lower-right corner. Thin dashed ellipses indicate the results after applying additional stellar mass cuts to each specific-SFR bin to equalize their mean stellar masses to that of the inactive population.
Grayscale symbols and error-bars indicate the mean cluster-centric radii and LOS velocity dispersions of cluster galaxies (including both star-forming and inactive sub-populations) split into four bins of stellar mass: $\log (\mathcal{M}/{\rm M}_{\odot}){<}10.4$ (light gray circle); $10.4{<}\log \mathcal{M}{<}10.6$ (gray triangle); $10.6{<}\log \mathcal{M}{<}10.85$ (mid-gray diamond); $\log \mathcal{M}{>}10.85$ (dark-gray square).}
\label{kinematics}
\end{figure}

To further delineate the connection between the kinematics of cluster galaxies and their ability to form stars, the sample of 24$\mu$m-detected cluster galaxies within $2.0\,r_{500}$ is split into four bins of specific-SFR and their mean cluster-centric radii ${\langle}r_{proj}/r_{500}{\rangle}$, and LOS velocity dispersions $\sigma_{los}({<}2\,r_{500})/\sigma_{\nu}$,  compared in Figure~\ref{kinematics}. 
Kinematic segregation of galaxies with diverse specific-SFRs is apparent. Galaxies with the highest specific-SFRs (${>}10^{-9.9}\,{\rm yr}^{-1}$; {\em blue cross}) have the highest velocity dispersion as a population, and a higher mean cluster-centric radius, than those star-forming galaxies with the lowest specific-SFRs (${<}10^{-10.5}\,{\rm yr}^{-1}$; {\em red cross}). There is a general progression towards lower mean radii and LOS velocity dispersions with decreasing specific-SFR. 
This progression continues to the inactive galaxy cluster galaxy population (not detected at 24$\mu$m; {\em black cross}), which has a significantly lower mean radius and velocity dispersion than any of the four sub-populations of star-forming galaxies.

The mean stellar masses of the star-forming cluster galaxies increases by 0.5\,dex from the highest specific-SFR bin to those in the lowest one, reflecting the systematic decline in specific-SFR with stellar mass for star-forming cluster galaxies \citep{haines13}. We may thus be concerned that these kinematical differences are in fact due to mass segregation rather than a sequence in declining specific-SFR. 

There is no evidence for mass segregation within our cluster galaxy population however, consistent with  \citet{vonderlinden}. Splitting the cluster galaxy population into bins of stellar mass ({\em grayscale symbols}), much less variation in the mean radii and LOS velocity dispersions is seen between stellar mass bins in comparison to those split by specific-SFR, and no overall trend with stellar mass is visible.
Moreover, the kinematic segregation by specific-SFR persists even if additional stellar mass cuts are applied to each specific-SFR bin to equalize their mean stellar masses ({\em thin dashed ellipses}).

\section{Mapping the continual accretion of galaxies onto massive clusters in the Millennium simulation}
\label{simulation}

To correctly interpret the previous observed trends in cluster galaxy properties, clusters must be placed in the cosmological context of continually accreting galaxies and groups from their surroundings. 
With this aim, we have examined the spatial distributions and orbits of galaxies in the vicinity of the 75 most massive clusters from the Millennium simulation \citep{springel_mill}, a cosmological dark matter simulation covering a ($500\,h^{-1}$Mpc)$^3$ volume. These clusters have present day virial masses in the range 4.0--23.$6{\times}10^{14}h^{-1}\,{\rm M}_{\odot}$, velocity dispersions of 630--1540\,km\,s$^{-1}$, and a median formation epoch, $z_{f}{=}0.59$. 
We have extracted $20{\times}20{\times}140h^{-3}\,{\rm Mpc}^{3}$ volumes centered on each cluster. 
These volumes are extended in the $z$-direction so that, for a distant observer viewing along this axis, all galaxies with LOS velocities within 5\,000\,km\,s$^{-1}$ of the cluster redshift are included, enabling projection effects to be fully account for and quantified. 

 There exists a full database of properties for each galaxy in the simulation, including positions, peculiar velocities, absolute magnitudes, stellar masses etc., based upon the {\sc galform} semi-analytic models (SAMs) of \citet{bower} at 63 snapshots throughout the life-time of the Universe to $z{=}0$. Similarly, another database provides the positions, velocities, masses ($M_{200}$) and radii ($r_{200}$) of each DM halo at each snapshot. For each galaxy and halo in a given snapshot, the database provides links to identify its most massive progenitor in the preceding snapshot, and so on, all the way back to its formation, allowing its mass growth and full merger history to be mapped in detail  (see e.g. De Lucia \& Blaizot 2007). This process also allows the orbit of each galaxy with respect to the cluster to be followed from formation to the present day, enabling us to determine its epoch of accretion ($z_{acc}$) into the cluster, defined here as the redshift at which the galaxy passes within $r_{200}(z)$ for the first time.

\begin{figure}
\centerline{\includegraphics[width=84mm]{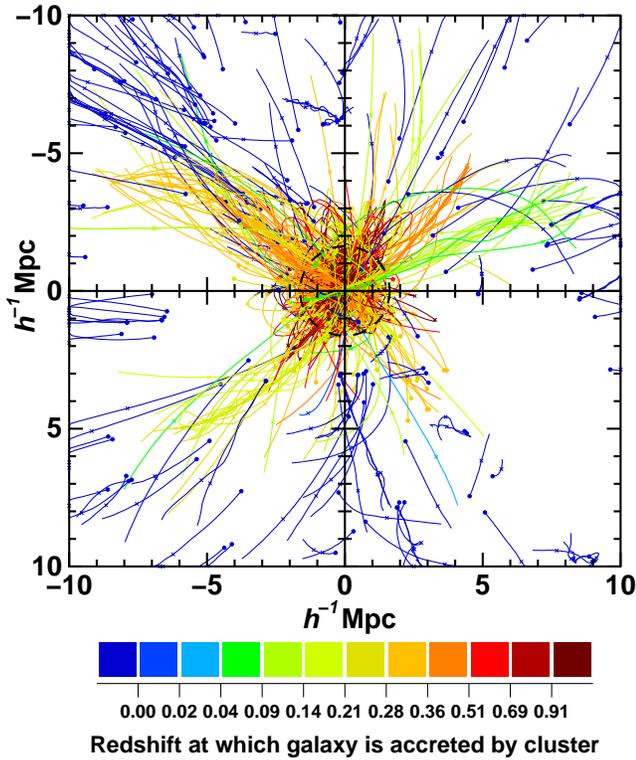}}
\caption{The orbits of galaxies about the 10th most massive cluster in the Millennium simulation, as viewed by a distant observer along the $z$-axis. The orbit of each galaxy with $\mathcal{M}{>}2{\times}10^{10}\,{\rm M}_{\odot}$ at $z{=}0$ is shown by a colored curve, tracing its movement from $z{=}0.76$ (snapshot 44) to $z{=}0.0$ (snapshot 63). The final location of the galaxy is marked by a dot, while the cross indicates its location at $z{=}0.41$ (snapshot 50). Each curve is color coded according to the epoch at which the galaxy is accreted into the cluster, as indicated at the bottom of the plot. This accretion epoch is defined as the snapshot at which the galaxy passes within $r_{200}(z)$ for the first time. Galaxies yet to pass within $r_{200}$ have mid-blue colors, while those accreted earliest into the cluster ($z_{acc}{>}0.51$) have red colors. The dashed black circle indicates the present day $r_{200}$ radius of the cluster. Only galaxies which would be spectroscopically identified as cluster members by the observer are shown.} 
\label{orbits}
\end{figure}

\begin{figure}
\centerline{\includegraphics[width=84mm]{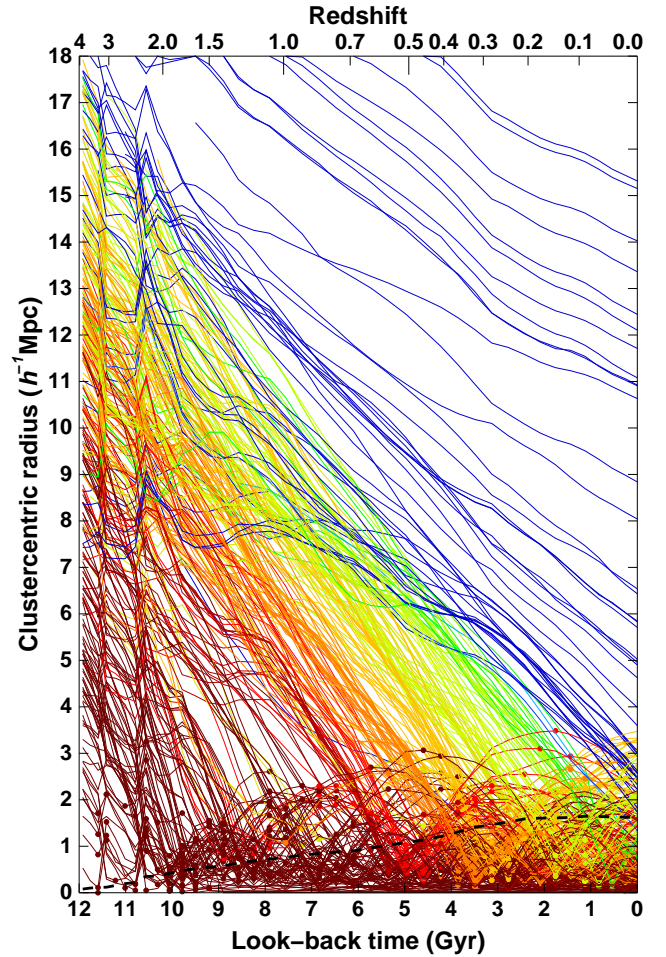}}
\caption{The infall of galaxies onto a massive cluster. Each galaxy with $\mathcal{M}{>}2{\times}10^{10}\,{\rm M}_{\odot}$ at $z{=}0$ is shown by a colored curve tracing its comoving cluster-centric distance as a function of look-back time. The curves are color coded according to the epoch at which the galaxy is accreted into the cluster, as in Figure~\ref{orbits}. The first pericenter and apocenter of each galaxy's orbit about the cluster are marked with dots. 
The black dashed curve indicates the evolution of the cluster radius $r_{200}$ with time.  
Only galaxies which would be spectroscopically identified as cluster members by the observer are shown.} 
\label{radial}
\end{figure}

\subsection{The infall of galaxies onto clusters}

Figure~\ref{orbits} shows the orbits of galaxies about the tenth most massive cluster ($M_{200}{=}9.9{\times}10^{14}h^{-1}\,{\rm M}_{\odot}$ at $z{=}0$) in the Millennium simulation, color coded according to accretion epoch.
Almost all the galaxies over the $20{\times}20\,h^{-2}\,{\rm Mpc}^{2}$ field-of-view are falling steadily into the cluster or have already been accreted. The complex large-scale structure around the cluster is apparent, including clear preferential directions for the galaxies to flow into the cluster, while other regions appear largely devoid of galaxies. Galaxies tend to be drawn first into the filaments from the surrounding field, then flow along the filaments into the cluster. 
While many galaxies are infalling as individual objects from the field, others are arriving into the cluster within galaxy groups (e.g. the tangle of green curves coming in from the right-hand side).

The full extent of the cluster's gravitational sphere of influence is revealed in Fig.~\ref{radial}, with all galaxies within $18\,h^{-1}\,$Mpc of the cluster falling steadily inwards. For all 75 clusters, the boundary between the infall regions and beyond, where galaxies remain attached to the Hubble flow, is found at a comoving distance ${\sim}1$0--2$0\,h^{-1}\,$Mpc from the cluster. 

The infall of galaxies into the cluster is highly coherent, at least for $z{\la}1$: the radial velocities of infalling galaxies at the same cluster-centric radius at a given epoch are all roughly the same, 
and the future trajectories and accretion epochs of an infalling galaxy can be accurately estimated simply on the basis of their current cluster-centric distance. For example, almost all galaxies which were $5\,h^{-1}\,$Mpc from the cluster 4 billion years ago, are due to be accreted into the cluster at $z{\sim}0.1$, as indicated by the parallel diagonal green colored curves. 

\begin{figure}
\centerline{\includegraphics[width=84mm]{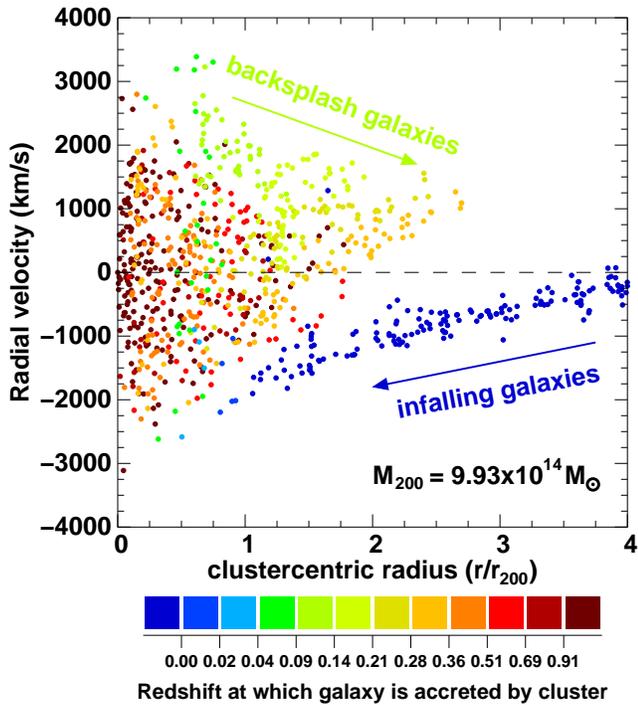}}
\caption{Radial phase-space diagram ($\nu_{radial}$ vs $r/r_{200}$) for galaxies with $\mathcal{M}{>}5{\times}10^{9}\,{\rm M}_{\odot}$ orbiting around the same massive cluster as in Figs~\ref{orbits} and~\ref{radial}. Galaxies are color coded according to their accretion epoch as indicated. 
} 
\label{phasespace}
\end{figure}

After accretion, galaxies remain bound to the cluster, but many have orbits which take them outside $r_{200}$ ({\em black dashed curve}), including some which bounce out as far as ${\sim}2.5\,r_{200}$. These galaxies, rebounding out of the cluster after their first pericenter passage are known as the ``back-splash'' population. 
The time-scale required for galaxies to reach pericenter after being accreted is of the order 0.5--0.8\,Gyr, while the orbital periods of the ``back-splash'' galaxies are much longer, only reaching their first apocenter 2--3\,Gyr after passing through the cluster for the first time. 

\subsection{Phase-space diagrams}

Figure~\ref{phasespace} shows the distribution of galaxies in radial phase-space: radial velocity ($\nu_{radial}$) versus cluster-centric radius ($r$) in units of $r_{200}$, for the same cluster,  
where $\nu_{radial}$ is the radial component of the galaxy's velocity relative to the cluster, including a component from the Hubble expansion ($H_{0}\,r$). 

The distribution of galaxies in phase-space splits into two reasonably well defined structures: a triangular-shaped virialized region containing galaxies which have passed through the cluster at least once; and a narrow stream of infalling galaxies ({\em blue points}) with negative radial velocities extending out to 4\,$r_{200}$ \citep{mamon, dunner}. As the infalling galaxies plunge into the deep gravitational potential well of the cluster, they are continually accelerated, reaching infall velocities of up to 3\,000\,km\,s$^{-1}$. 
After passing through pericenter, these galaxies reappear along the top edge of the triangular region as back-splash galaxies, coherently progressing outwards and slowing down with increasing $z_{acc}$ towards the right-hand apex at ${\ga}2\,r_{200}$ that marks the apocenter of the orbits of those galaxies accreted ${\sim}3$\,Gyr ago.  
The radial phase-space diagram retains much of the information regarding the epoch of accretion of a galaxy, allowing this epoch to be accurately estimated for galaxies based on their location in the diagram, at least for those accreted within the last 3\,Gyr. Only those which were accreted much earlier have had time for their orbits to become mixed. 

\begin{figure}
\centerline{\includegraphics[width=84mm]{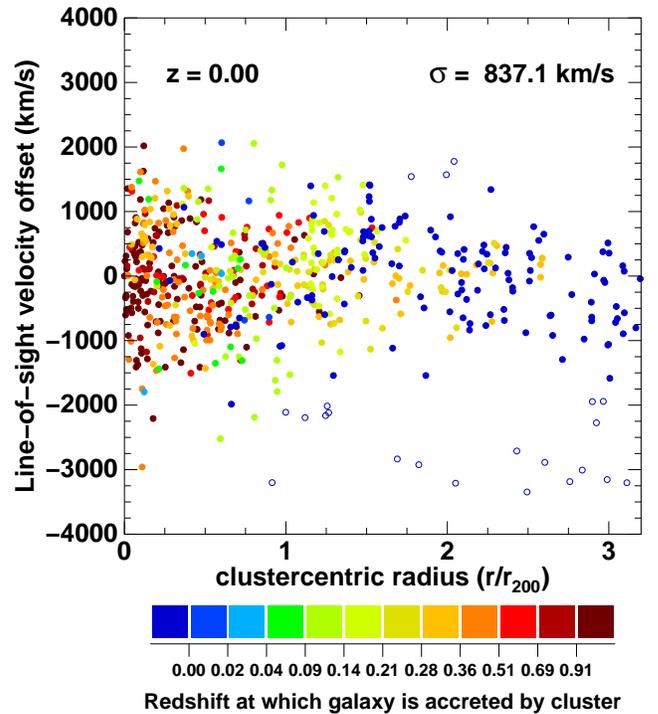}}
\caption{The caustic diagram ($\Delta \nu_{los}$ vs $r_{proj}/r_{200}$) for $\mathcal{M}{>}2{\times}10^{10}\,{\rm M}_{\odot}$ galaxies orbiting around the same massive cluster, as viewed by distant observer along the $z$-axis. Galaxies are color coded according to their accretion epoch as in Figs.~\ref{orbits}--\ref{phasespace}. Galaxies within  $20\,h^{-1}$\,Mpc of the cluster center at $z{=}0.0$ are indicated by solid symbols, while those at comoving cluster-centric radii $r{>}20\,h^{-1}$\,Mpc are shown by open symbols. The velocity dispersion of galaxies within a projected cluster-centric radius $r_{proj}{<}r_{200}$ is indicated.
} 
\label{zspace}
\end{figure}

Figure~\ref{zspace} shows the observable counterpart to Fig.~\ref{phasespace}: the caustic diagram, which plots the LOS velocity of galaxies relative to the cluster redshift ($\Delta\nu_{los}$) against projected cluster-centric radius ($r_{proj}/r_{200}$) for the same cluster, as viewed by a distant observer along the $z$-axis. The relative LOS velocity of galaxies combines the LOS component of their peculiar velocities with the contribution to their redshifts from the Hubble expansion: $\Delta \nu_{los}{=} (\nu_{z,gal} - \nu_{z,cl}) + H_{0}\,d_{z}$, where $\nu_{z,gal}$ and $\nu_{z,cl}$ are the LOS peculiar velocities of the galaxy and cluster DM halo respectively, and $d_{z}$ is the distance between the galaxy and the cluster halo along the line of sight. 

\begin{figure*}
\centerline{\includegraphics[width=178mm]{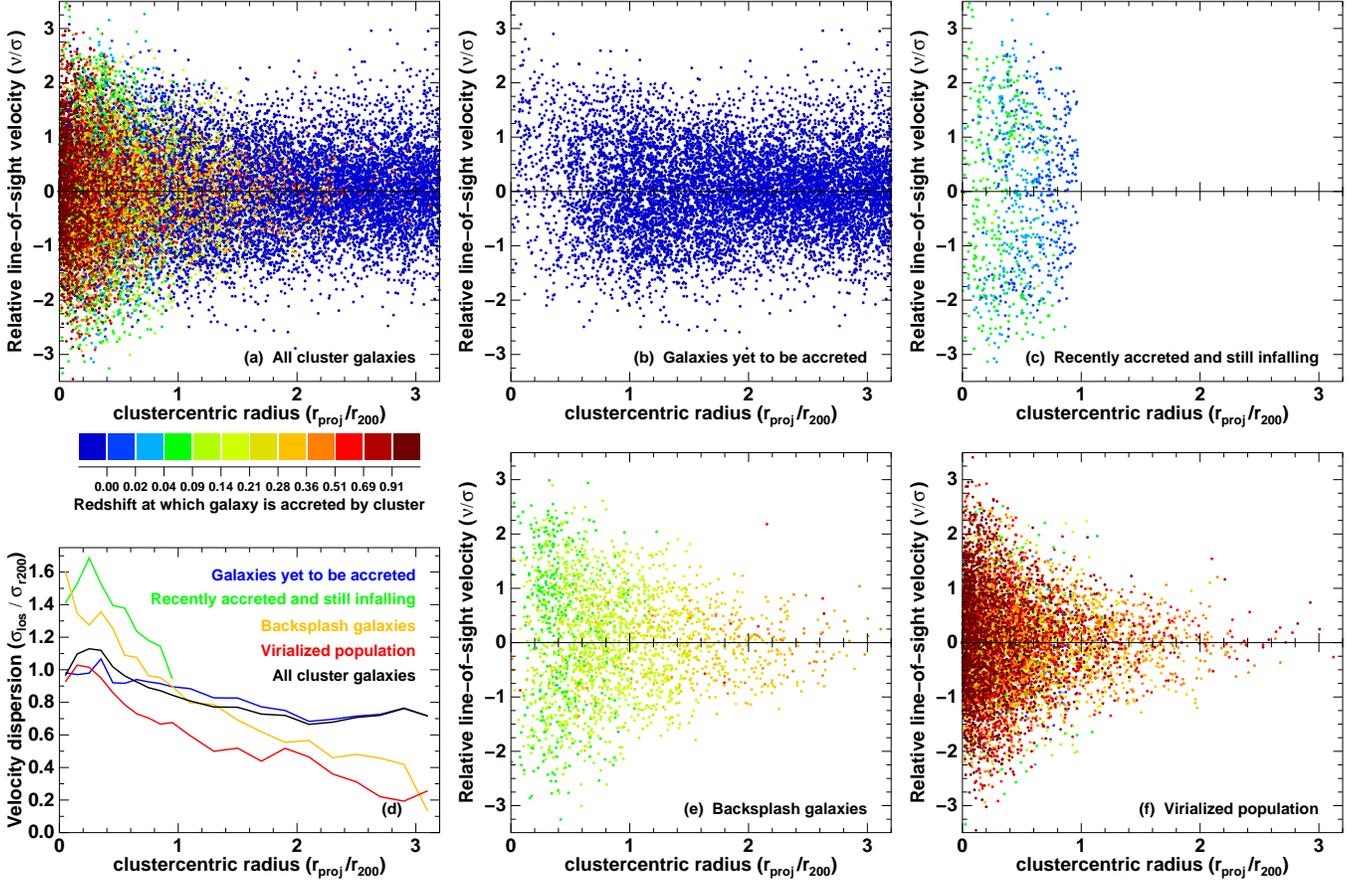}}
 \caption{Stacked phase-space diagram ($\Delta \nu_{los}/\sigma_{\nu}$ vs. $r_{proj}/r_{200}$) for the 75 most massive clusters in the Millennium simulation at $z{=}0.0$ (panel a). Each point indicates an $\mathcal{M}{>}2{\times}10^{10}\,{\rm M}_{\odot}$ galaxy from the \citet{bower} semi-analytic model catalog, colored according to when it was accreted onto the cluster. The four panels on the right split these cluster galaxies into dynamical sub-populations: (b) just those galaxies yet to pass within $r_{200}$ and be accreted into the cluster; (c) infalling galaxies ($\nu_{radial}{<}0$) which have passed within $r_{200}$, but have yet to reach pericenter; (e) back-splash galaxies which have passed through pericenter and are on their way back out of the cluster ($\nu_{radial}{>}0$) and are yet to reach apocenter; (f) the virialized cluster population which have passed through apocenter. The lower left panel (d) displays the velocity dispersion profiles ($\sigma_{los}(r)/\sigma_{\nu}$) of each sub-population as well as that of the overall cluster population.}
\label{zspace_stacked}
\end{figure*}

 The distribution of cluster galaxies shows the typical ``trumpet''-shaped caustic profile. Throughout this paper we refer to ``spectroscopic'' cluster members as those galaxies which lie within the caustic profiles determined for each of the 75 clusters. This is not the same as galaxies which are identified as satellites of the cluster DM halo, which are those physically located within the cluster halo ($r{<}r_{200}$), or even the population of accreted galaxies that have passed within $r_{200}(z)$ at some point in their history (but which may now be outside $r_{200}$). 
For this cluster, the separation between those galaxies that would be identified as ``spectroscopic'' cluster members, and those which are clearly background objects, roughly corresponds to a separation between those physically within $20h^{-1}$\,Mpc of the cluster center ({\em solid points}), and those beyond this radius ({\em open points}).
Indeed, all the galaxies shown earlier in Figure~\ref{radial} are ``spectroscopic'' cluster members within a projected cluster-centric radius $2.0\,r_{200}$. Examining each of the 75 clusters individually, the cluster-centric distances that best separate ``spectroscopic'' members and clear fore/background objects are $19{\pm}3\,h^{-1}\,$Mpc ($13{\pm}2\,r_{200}$). 

Unlike the radial phase-space diagram, it is not possible to select individual galaxies from a specific region of the caustic diagram and then identify it as an infalling, recently accreted or virialized galaxy. However, several trends within the distribution of galaxies in the caustic diagram can be seen. 
First, those galaxies accreted earliest ($z_{acc}{\ga}0.4$; {\em red points}) are spatially localized in the cluster core with typical LOS velocities ${\la}1$\,000\,km\,s$^{-1}$. 
Second, beyond $r_{proj}{\ga}1.8\,r_{200}$ the bulk of galaxies have yet to be accreted into the cluster while the remainder all appear to be back-splash galaxies accreted ${\sim}3$\,Gyr ago. Finally, many of the galaxies with the largest LOS velocities ($\ga1$\,000\,km\,s$^{-1}$) have only recently arrived into the cluster ({\em green points}). 
Although these general trends hold for the vast majority of the clusters in our sample, there is significant cluster-to-cluster variation in the spatial distributions and relative contributions of galaxies accreted at different epochs, as expected given their dynamical immaturity.

\subsection{Stacking the clusters}

To account for the cluster-to-cluster scatter in a statistical way, the caustic diagrams for all 75 clusters are stacked to produce Fig.~\ref{zspace_stacked}a. 
The cluster-centric radius of each cluster member is scaled by the $r_{200}$ of that cluster, and the LOS velocities are scaled in units of $\sigma_{\nu}(r_{proj}{<}r_{200})$, the LOS velocity dispersion of all ``spectroscopic'' members within $r_{200}$. 

To demonstrate how the distribution of galaxy populations in the caustic plot provides information about their dynamical evolution and accretion history, the right-hand panels show the caustic diagrams after splitting the cluster population into four dynamical sub-populations, while panel (d) plots each of their velocity dispersion profiles $\sigma_{los}(r)/\sigma_{\nu}$ alongside that for the overall cluster population. 
First, galaxies yet to pass within $r_{200}$ and be accreted (panel b) are found at all cluster-centric radii, becoming increasingly dominant with radius. The fall in numbers towards the cluster core ($r_{proj}{\la}0.6\,r_{200}$) is due to the area of sky covered in any given narrow radial slice scaling as $r$, rather than a physical decline in the surface density of such galaxies at low radii (see Fig.~\ref{profile}). The velocity dispersion of these galaxies remains relatively constant with radius, as they have yet to approach the cluster core 

Panel (c) shows those infalling galaxies which have recently passed within $r_{200}$ for the first time, but have yet to reach pericenter. These objects are all found within $r_{200}$ by default. They have the highest LOS velocities of all our sub-populations ($\sigma_{los}(r){\ga}1.4\,\sigma_{\nu}$), being fully accelerated as they fall deep into the gravitational potential well of the cluster core, and at late epochs when the clusters are much more massive than at any previous point in their history. 

\begin{figure*}
\centerline{\includegraphics[width=178mm]{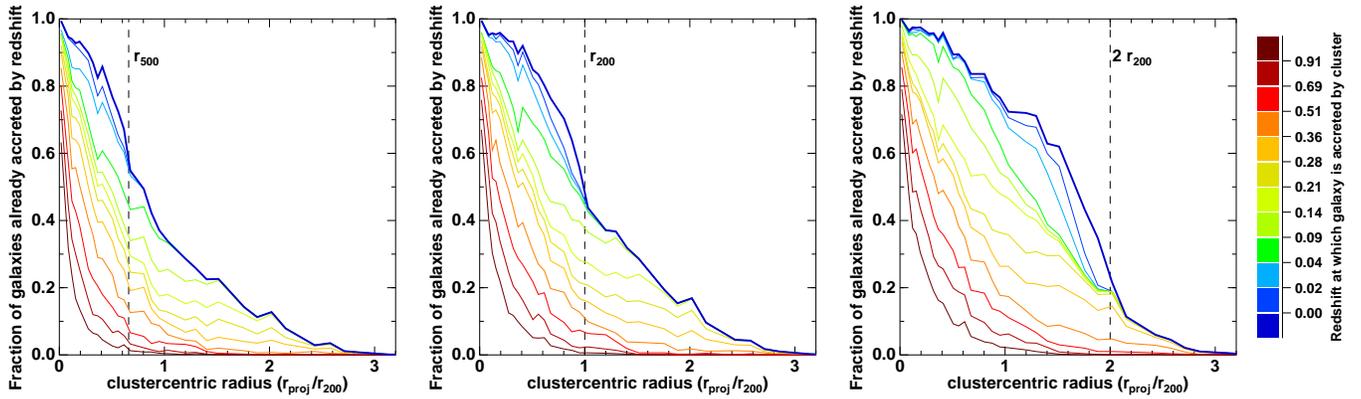}}
\caption{Radial population gradients in clusters. Each curve shows the fraction of ``spectroscopic'' cluster galaxies which have been accreted by a given redshift as a function of projected cluster-centric radius (normalized by $r_{200}$) averaged over our stacked sample of 75 massive clusters observed at $z{=}0.0$, color coded according to accretion redshift as indicated on the far right of the plot. The three panels indicate the trends produced when considering three definitions for describing when a galaxy is accreted into the cluster, taken to be the snapshot when the galaxy passes within $r_{500}$ ({\em left panel}), $r_{200}$ ({\em middle panel}) or $2\,r_{200}$ ({\em right panel}). The vertical dashed line in each panel indicates the corresponding radius at which the accretion epoch for the curves in that panel are defined.
} 
\label{gradients}
\end{figure*}

Panel (e) shows the back-splash population of cluster galaxies which have passed through pericenter, and are now heading back out away from the cluster center ($\nu_{radial}{>}0$), but have yet to reach apocenter. These galaxies were typically accreted 1--4\,Gyr ago. This population shows a triangular distribution in the caustic diagram, with high LOS-velocities ($\sigma_{los}{\sim}1$.3--1.$5\,\sigma_{\nu}$) in the cluster core, having recently completed their first infall,  which steadily fall as the galaxies rebound out of the cluster, slowing down as they attempt to climb back out of its potential well. The back-splash population extends as far out as 2--3\,$r_{200}$, where they can be differentiated from the infalling population by their characteristically low LOS-velocity dispersions ($\sigma_{los}(r){\sim}0$.4--0.$55\,\sigma_{\nu}$), some ${\sim}3$5\% lower than the overall cluster population at the same cluster-centric distances \citep[see][]{gill}. 

Finally, the virialized population (panel f) are those galaxies which were accreted at early epochs ($z_{acc}{\ga}0.4$) and have all passed through apocenter of their first orbit. This population also presents a triangular distribution in the caustic diagram, but is much more concentrated towards the cluster core (particularly at ${\la}0.2\,r_{200}$) than the back-splash galaxies, and has lower LOS-velocity dispersions at all radii. 
We identify this population with those galaxies that either formed locally or were accreted when the cluster's core was being assembled. Their low velocities reflect the fact that the system they fell into was much less massive than the present-day cluster. 

\subsection{Radial population gradients}
\label{sim:radial}

Figure~\ref{gradients} shows the radial population gradients obtained when stacking the 75 massive clusters as observed at $z{=}0.0$. 
The thick blue curve in the middle panel shows how the fraction of ``spectroscopic'' cluster members 
which have been accreted varies as a function of projected cluster-centric radius, $r_{proj}/r_{200}$. This fraction drops slowly from 100\% at the cluster core to ${\sim}9$0\% at $0.4\,r_{200}$, before falling at an ever increasing rate down to ${\sim}5$0\% by $r_{200}$, and approaching zero by ${\sim}3\,r_{200}$. For $r_{proj}{>}r_{200}$, this contribution represents the ``back-splash'' population galaxies which have previously been accreted into the cluster, but now have bounced back out beyond $r_{200}$.

The remaining curves show the effects on this fraction by progressively excluding galaxies which were accreted after a given redshift. 
The first three curves ($z_{acc}{<}0.09$) only diverge from the top curve at $r_{proj}{<}r_{200}$, as the most recently accreted galaxies haven't had sufficient time to pass though the cluster and go back out beyond $r_{200}$. 
As those galaxies accreted 1--3\,Gyr ago are progressively excluded, the curves become increasingly steep within $r_{200}$, and flatter beyond $r_{200}$. The bulk of the ``back-splash'' population found beyond $r_{200}$ were accreted between $z{=}0.14$ and $z{=}0.51$.

The remaining panels show the effect of changing the nominal radius for identifying when a galaxy has been accreted into a cluster. Reducing the accretion radius from $r_{200}$ to $r_{500}$ ({\em left panel}) simply squeezes the curves inwards by a comparable amount. Pushing the accretion radius outwards to $2\,r_{200}$ ({\em right panel}) increases the prominence of the bump produced by recently accreted galaxies, but greatly reduces the contribution from back-splash galaxies, as so few galaxies which pass within $2\,r_{200}$ rebound from the cluster beyond this radius.

The key aspect of all these curves is that irrespective of accretion epoch, $z_{acc}$, or the precise radius used to define accretion (within reasonable limits), the resulting radial population gradient drops from ${\sim}1$00\% in the cluster core to zero by ${\sim}3\,r_{200}$. This steep gradient is primarily due to the complementary increase with radius in the fraction of ``interloper'' galaxies that have yet to be accreted into the cluster among the ``spectroscopic'' cluster population, from zero in the cluster core to 100\% by ${\sim}3\,r_{200}$. 
A second contribution to the steepness comes from the correlations between the epoch of accretion of {\em satellite} galaxies and their present cluster-centric distance \citep{delucia}, with those satellite galaxies close to the cluster core having been accreted significantly earlier on average than those located close to the virial radius (both physically or in projection). 
\citet{gao04} find the same radial trend for sub-halos, with the median accretion redshift of sub-halos in massive cluster halos decreasing from $z_{acc}{\sim}1.0$ in the cluster core to $z_{acc}{\sim}0.4$ at $r_{200}$ (their Fig. 15).

\section{Constraining star formation quenching models by comparison to the observed SF--radius trends}
\label{sec:quenching}

Figures~\ref{sf_radius}--\ref{UV_sf_radius} showed that the fraction of massive cluster galaxies with ${\rm SFR}_{IR}{>}2.0\,{\rm M}_{\odot}\,{\rm yr}^{-1}$ or blue UV--optical colors increases steadily with cluster-centric radius, but at the largest radii probed, the $f_{SF}$ remained significantly (20--30\%) below that seen in coeval field galaxies in both cases.

\begin{figure}
\centerline{\includegraphics[width=84mm]{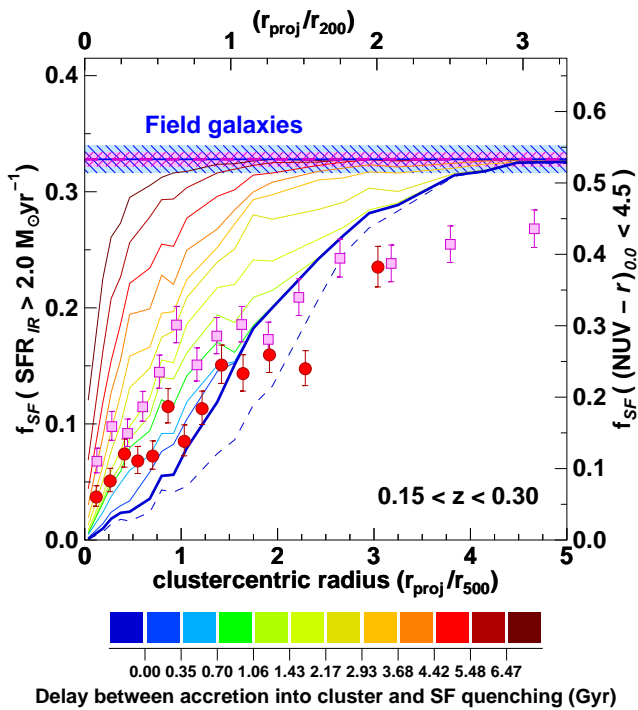}}
\caption{Comparison of the observed radial population gradients with predictions from cosmological simulations. The red points show the fractions of $\mathcal{M}{>}2{\times}10^{10}\,{\rm M}_{\odot}$ cluster galaxies with obscured ${\rm SFR}{>}2.0\,{\rm M}_{\odot}\,{\rm yr}^{-1}$ from Fig.~\ref{sf_radius}, while the magenta squares show the corresponding SF-radius relation for UV-selected star-forming galaxies (Fig.~\ref{UV_sf_radius}). 
The blue horizontal line indicates the corresponding fraction of field galaxies ($\mathcal{M}{>}2{\times}10^{10}\,{\rm M}_{\odot}$; $0.15{<}z{<}0.30$) with ${\rm SFR}_{IR} > 2.0\,{\rm M}_{\odot}\,{\rm yr}^{-1}$, while the blue shaded region indicates the $1\sigma$ confidence levels. 
The thick blue diagonal curve shows the predicted SF--radius relation obtained from our ``observations'' of the 75 massive clusters in the Millennium simulation, assuming that infalling galaxies have the same $f_{SF}$ as our observed field galaxy sample, and star-formation is immediately quenched upon being accreted into the cluster, i.e. passing within $r_{200}$ for the first time.  The colored curves show the effects of delaying this quenching by a time $\Delta t$ after the galaxy is accreted into the cluster, as indicated by the color scale below the plot.
The dashed blue curve shows the effect on the predicted SF--radius relation of changing the cluster-centric radius at which quenching is initiated from $r_{200}$ out to $2\,r_{200}$. 
} 
\label{sf_radius_models}
\end{figure}

In Figure~\ref{sf_radius_models} we attempt to reproduce these two SF-radius trends by a simple toy model in which the star formation of infalling field galaxies is instantaneously quenched at the moment they pass within $r_{200}$ of the cluster for the first time, or after a certain time delay ($\Delta t$). The fraction of star-forming galaxies among this infalling field population is set to match our observed coeval field galaxy sample. The stacked radial population gradients for $\mathcal{M}{>}2{\times}10^{10}\,{\rm M}_{\odot}$ galaxies from the same 75 massive clusters in the Millennium simulation are reproduced, as they would appear if observed at $z{=}0.21$, to best match the redshifts of the LoCuSS sample. The model galaxy positions and velocities relative to the cluster halo are now measured as they stood at $z{=}0.21$, while the clusters are stacked using their $r_{200}$ and $\sigma_{\nu}$ values measured at $z{=}0.21$. At $z{=}0.21$ these clusters have M$_{200}$ masses in the range 2.6--21.$7{\times}10^{14}h^{-1}\,{\rm M}_{\odot}$, with a median $M_{200}$ of $5.0{\times}10^{14}h^{-1}\,{\rm M}_{\odot}$.

The predicted SF--radius relation in the case that star formation in all infalling galaxies is instantaneously quenched upon accretion ({\em thick blue diagonal curve}) is qualitatively similar in form to the observed trends, and consistent with the data points at ${\sim}1-2\,r_{500}$, suggesting that this is to first order a reasonable assumption, as found previously by \citet{balogh00} and \citet{haines09a}. The model radial gradient is too steep however, resulting in predicted values of $f_{SF}$ that are much higher than our data points in the range 1.8--3.$0\,r_{500}$, and too low in the cluster core ($r_{proj}{\la}0.8\,r_{500}$). 

The remaining colored curves show the effects of delaying the moment at which quenching occurs, by terminating star-formation only in those galaxies accreted into the cluster more than $\Delta t$\,Gyr prior to observation, corresponding to the ``delayed-then-rapid'' quenching scenario of \citet{wetzel13}. The ``excess'' obscured star-formation ({\em red points}) observed in the cluster core can then be reasonably reproduced by a model with a short quenching delay of the order  0.3--1.0\,Gyr  ({\em light-blue/green curves}). Much longer quenching time-delays ($\Delta t{\ga}3$\,Gyr) are clearly excluded, as they leave too many star-forming galaxies at $r_{proj}{\sim}1-2\,r_{500}$. 

One possible way to reconcile the model predictions with our data at $\sim$3--5$\,r_{500}$ would be to initiate the quenching process at larger radii. 
The dashed curve shows the radial population gradient produced when star-formation in all infalling galaxies is immediately quenched when they pass within $2\,r_{200}$ for the first time (rather than $r_{200}$ as before). 
While the fraction of star-forming galaxies has now been reduced sufficiently at ${\sim}2\,r_{500}$, the gradient of the model trend remains much steeper than that observed, and vastly under-predicts the fraction of star-forming galaxies at $r_{proj}{\la}1.5\,r_{500}$.

\begin{figure}
\centerline{\includegraphics[width=84mm]{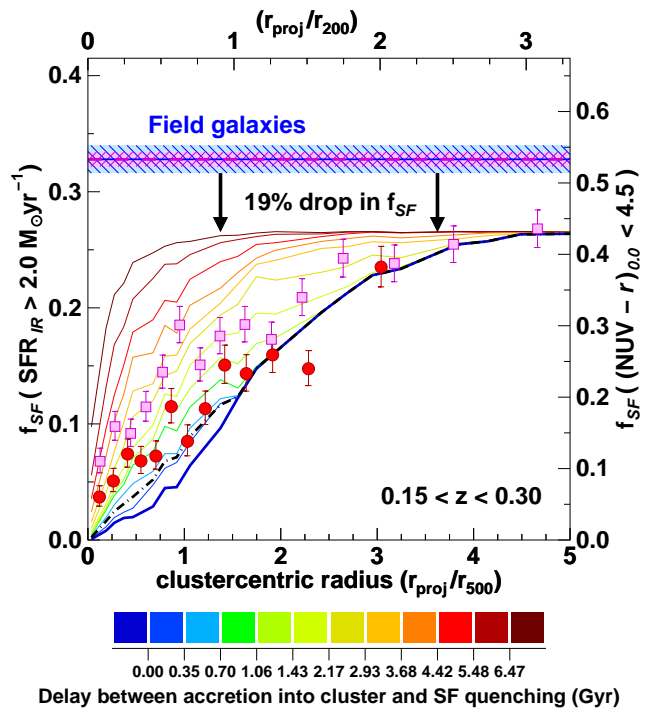}}
\caption{Radial population gradients. The red and magenta points show the observed IR- and UV-based SF--radius relations as in Figure~\ref{sf_radius_models}. The colored curves are the same as those in Figure~\ref{sf_radius_models}, except that they now assume that the infalling galaxies have a fraction $f_{SF}$ which has been reduced by 19\% with respect to that seen in the general field (blue horizontal line). The black dot-dashed curve indicates the predicted SF--radius relation in the case that star formation is instantaneously quenched when galaxies reach pericenter.
}
\label{sf_lower}
\end{figure}

The model curves provide little or no leeway to reproduce the low $f_{SF}$ observed over 3--5$\,r_{500}$ in the NUV-based SF--radius relation ({\em magenta squares}), as they all approach the field value at 3.5--4.0\,$r_{500}$ (${\sim}2$.5--3.$0\,r_{200}$), the radius at which the fraction of back-splash galaxies falls to zero (Fig.~\ref{gradients}), irrespective of the radius at which quenching is initiated. 

The only way to improve the model fits to these observations would be to allow the fraction of star-forming galaxies among the infalling population to be {\em lower} than that observed among the general coeval field population. 
We consider the simplest approach in Figure~\ref{sf_lower}, that of simply reducing the $f_{SF}$ of cluster galaxies by a single fixed amount (19\%) from the value observed in the field, at all radii, to model the impact of whatever physical process is reducing star formation among galaxies in the infall regions of clusters. One feasible mechanism to achieve this is via the ``pre-processing'' of galaxies within galaxy groups which are subsequently accreted into clusters. The cosmological simulations of \citet{gabor} suggest that ${\sim}4$0\% of satellite galaxies within clusters are pre-processed, the fraction decreasing weakly with cluster-centric radius.  
The overall IR-based SF--radius relation can now be reproduced at all radii, 
via a model in which star-formation is quenched in galaxies ${\sim}0$.7--1.5\,Gyr after being accreted into the cluster. This occurs on average {\em after} the galaxy has passed through the cluster core, as the observed trend lies above that predicted for the case in which star formation is quenched in galaxies at the moment they reach pericenter ({\em black dot-dashed curve}).

The NUV-based SF--radius relation is systematically above that of the IR-based inside ${\sim}2.5\,r_{500}$, 
and is best-fit by a model is  which star-formation is quenched in galaxies ${\sim}2$.1--3.6\,Gyr after accretion, or slightly later than that suggested by the IR-based relation. 

For all these model curves, to use the terminology of \citet{peng,peng12}, we assume that the environmental quenching process is 100\% efficient ($\epsilon_{sat}{=}1$), i.e. {\em all} star-forming galaxies are quenched $\Delta t$\,Gyr after accretion into the cluster. 
The result of decreasing $\epsilon_{sat}$ for clusters would leave the model curves unchanged at large radii ($r_{proj}{\ga}3\,r_{200}$), but squeeze the curves upwards in the core, effectively reducing the radial population {\em gradient} by a factor $(1-\epsilon_{sat})$. The fact that the observed gradients are steep, with $f_{SF}$ increasing fivefold over 0--3\,$r_{200}$, necessitates a high environmental quenching efficiency, $\epsilon_{sat}{\ga}0.8$, otherwise there would be evidence of a residual population of primordial cluster galaxies with ongoing star-formation in cluster cores ($r_{proj}{\la}0.1\,r_{200}$). 

\subsection{Uncertainties in the model trends}

One concern about attempting to fit the SF--radius relation with model curves, is that the latter require the semi-analytic models to reliably follow the orbits of cluster galaxies long after their accretion, and accurately model the long-term evolution of their stellar masses once they become satellites. The Millennium simulation itself considers only the dark matter component, the baryons bolted on afterwards via SAMs. 

When galaxies are accreted into massive clusters, their parent DM halos become sub-halos of the cluster halo, losing mass continuously through tidal stripping, until in many cases they fall below the resolution limit of the simulation, dissolving into the parent halo or are completely disrupted \citep{gao04,weinmann10}.  
The galaxies hosted by these sub-halos are much more compact and tightly bound than the dark matter, and as long as some of the surrounding sub-halo survives, are not expected to suffer significant stellar mass loss, although their diffuse hot gas halo is likely to be lost at a rate commensurate with that of the parent sub-halo \citep{guo11}. 
Prior to their accretion into the clusters, model galaxies at our lower stellar mass limit of $2{\times}10^{10}{\rm M}_{\odot}$ typically have parent DM halos of masses ${\sim}6{\times}10^{11}h^{-1}{\rm M}_{\odot}$ in the Millennium simulation, comprising ${\sim}730$ DM particles, and hence must suffer $>95$\% stripping before their parent sub-halo falls below the mass resolution limit of 20 particles \citep{springel_mill}, a process which typically takes ${\sim}5$\,Gyr \citep{weinmann10}. 
Even after their parent sub-halo has been entirely disrupted, they are expected to survive as ``orphan'' galaxies, although now they will be likely subject to significant stellar mass loss via tidal stripping and may be completely disrupted. 

The inclusion and treatment of ``orphan'' galaxies by SAMs is required to explain the cuspy radial satellite number density profiles observed within cluster halos (Fig.~\ref{profile}), which are much steeper than the relatively flat radial distributions of DM sub-halos within $0.3\,r_{500}$ \citep{budzynski,vogelsberger}. These SAMs resort to {\em ad hoc} prescriptions for mass-stripping and adjustment of the orbits of these ``orphan'' galaxies within the cluster halos, resulting in strong variations in the radial satellite number density profiles within cluster cores \citep{budzynski}, and the fractions of ``orphans'' in the cluster satellite population, from 25\% in the \citet{guo11} model to 50\% in the \citet{bower} model \citep{gifford}. This likely explains the inability of the predicted radial profile of model cluster galaxies from the \citet{bower} SAM to match the observed radial profile at $r_{proj}{\la}0.1\,r_{500}$, overestimating the number density of cluster galaxies by a factor 2--3 (Fig.~\ref{log_profile}). 

 We should therefore be cautious about using model curves, such as those in Figs.~\ref{sf_radius_models} and~\ref{sf_lower}, that depend upon the radial distribution and numbers of those galaxies accreted earliest into the clusters, and which have suffered repeated interactions over multiple orbits within the ever growing cluster halo. All the model curves in these two figures do this, even those referring to galaxies yet to be accreted into cluster, as they depend upon the relative contributions of those accreted after a given epoch, with all those accreted before the same epoch.

\subsection{Radial galaxy surface density profiles}
\label{sec:profiles}
 
\begin{figure}
\centerline{\includegraphics[width=84mm]{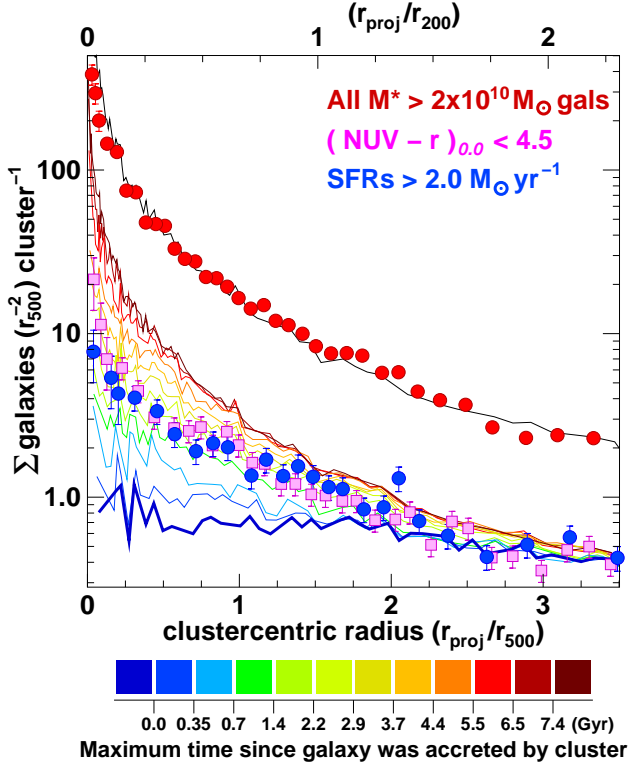}}
\caption{Blue symbols show the $\Sigma(r)$ profile of $\mathcal{M}{>}2{\times}10^{10}\,{\rm M}_{\odot}$ galaxies with obscured star formation at rates SFR$_{IR}{>}2.0\,{\rm M}_{\odot}{\rm yr}^{-1}$ averaged over our 30 clusters. Magenta squares show the $\Sigma(r)$ profile for unobscured star-forming galaxies with \mbox{$({\rm NUV}{-}r)_{0.0}{<}4.5$}, normalized to match the profile of obscured star-forming galaxies at $r_{proj}{\ga}1.5\,r_{500}$. 
The thick blue curve shows the predicted surface density profile of infalling galaxies yet to pass within $r_{200}$, obtained from our ``observations'' of the 75 massive clusters in the Millennium simulation at $z{=}0.21$, normalized to fit the observed radial profiles of star-forming galaxies at large radii. The remaining colored curves show the predicted radial profiles produced by including also those galaxies accreted into the clusters within the last $\Delta t$\,Gyr as indicated by the color scale. The black curve shows the predicted surface density profile for all ``spectroscopic'' member galaxies of the same 75 clusters, scaled to best match the observed ensemble $\Sigma(r)$ profile of {\em all} $\mathcal{M}{>}2{\times}10^{10}\,{\rm M}_{\odot}$ cluster galaxies ({\em red points}).  
}
\label{profile}
\end{figure}

Figure~\ref{profile} compares the radial galaxy surface density profiles, $\Sigma(r)$, of star-forming galaxies with model surface density profiles considering just those galaxies accreted within the last $\Delta t$ Gyr or which have not yet been accreted. This plot largely resolves the above issues, by focusing solely on star-forming galaxies, which are most likely to still have surviving parent sub-halos, and comparison model curves that contain just the most recent arrivals into clusters, and hence minimizing the uncertain contribution from ``orphan'' galaxies. 

The surface density of star-forming (${\rm SFR}_{IR}{>}2.0\,{\rm M}_{\odot}{\rm yr}^{-1}$) cluster galaxies from our ensemble of 30 clusters ({\em blue points}) declines steadily with radius 
out to ${\sim}3\,r_{500}$, with no evidence of flattening off inside $r_{500}$.  
This immediately rules out models in which star formation is instantaneously quenched when galaxies are accreted into clusters ({\em thick blue curve}), as these produce radial profiles which are essentially flat within $2\,r_{500}$. As recently accreted galaxies are progressively included, the radial density profile steadily builds up and steepens within $2\,r_{500}$. The best fitting curve to the {\em Spitzer} data is obtained by considering those model cluster galaxies accreted within the last $\Delta t {=} 2.1_{-0.7}^{+0.8}$\,Gyr, with a $\chi^{2}$ value of 21.15 for 23 data points and two degrees of freedom ($\phi$ and $\Delta t$). The uncertainties in $\Delta t$ are derived as the values for which the $\chi^{2}$ value has increased by 2.30 from the minimum value. 

 Outside of $0.3\,r_{500}$, the shape of the $\Sigma(r)$ profile for unobscured star-forming galaxies ({\em magenta squares}; $\mathcal{M}{>}2{\times}10^{10}\,{\rm M}_{\odot}$, $(NUV-r)_{0.0}<4.5$), coincides well with that of the obscured star-forming population.  
In contrast, the $\Sigma(r)$ profile for the UV-selected star-forming galaxies steepens more rapidly in the cluster core to form a cusp, paralleling that seen for the overall cluster population ({\em red points}). The best fitting curve to the NUV data has $\Delta t {=} 3.2{\pm}0.4$\,Gyr ($\chi^{2}{=}29.57$ for 32 data points), implying that the NUV emission takes 1.1\,Gyr longer after accretion to be shut down than the 24$\mu$m emission. 

\begin{figure}
\centerline{\includegraphics[width=84mm]{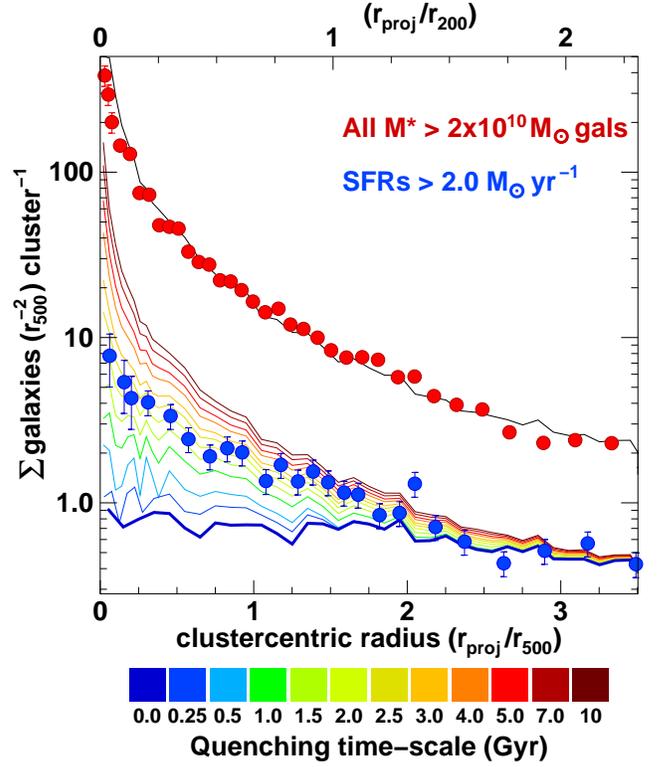}}
\caption{Blue symbols show the $\Sigma(r)$ profile of $\mathcal{M}{>}2{\times}10^{10}\,{\rm M}{\odot}$ galaxies with obscured star formation at rates SFR$_{IR}{>}2.0\,{\rm M}_{\odot}{\rm yr}^{-1}$. Red symbols indicate the corresponding $\Sigma(r)$ profile for {\em all} $\mathcal{M}{>}2{\times}10^{10}\,{\rm M}{\odot}$ galaxies. The thick blue curve shows the predicted surface density profile of infalling galaxies yet to pass within $r_{200}$, normalized to fit the observed surface density profile of star-forming galaxies at large radii. The remaining colored curves show the predicted surface density profiles for star-forming galaxies accreted into the clusters and subsequently have their star-formation rates decline exponentially with a range of quenching time-scales ($t_{Q}{=}0$.25--10\,Gyr). 
}
\label{profile_exp}
\end{figure}
 
 Figure~\ref{profile_exp} replots the $\Sigma(r)$ profile for obscured star-forming cluster galaxies ({\em blue points}), but now compares it with slow quenching models. In this scenario model star-forming galaxies are accreted into the cluster and subsequently gradually quenched, their SFRs declining exponentially on a quenching time-scale $t_{Q}$ until their SFRs fall below our nominal limit of 2.0\,M$_{\odot}$\,yr$^{-1}$. These model star-forming galaxies are given initial SFRs taken at random from our observed sample of coeval {\em field} star-forming galaxies (${\rm SFR}{>}2\,{\rm M}_{\odot}\,{\rm yr}^{-1}$; $0.15{<}z{<}0.30$),  the SFR distribution of which is shown in Fig.~2 of \citet{haines13}.
This should be reasonable given that the infrared luminosity functions of cluster and field galaxies are indistinguishable \citep{finn10,haines11a,haines11b,haines13}, and the specific-SFRs of star-forming galaxies in infall regions are indistinguishable from those in coeval field samples \citep{haines13}.
The colored curves show the predicted surface density profiles for quenching time-scales $t_{Q}$ in the range 0.25--10.0\,Gyr. The best fit model to observations has a quenching time-scale $t_{Q}{=}2.19{\pm}0.41$\,Gyr, with a $\chi^{2}$ value of 20.80 for 23 data points  and two degrees of freedom ($\phi$, $t_{Q}$). 

\subsection{Dynamical analysis}

\begin{figure}
\centerline{\includegraphics[width=84mm]{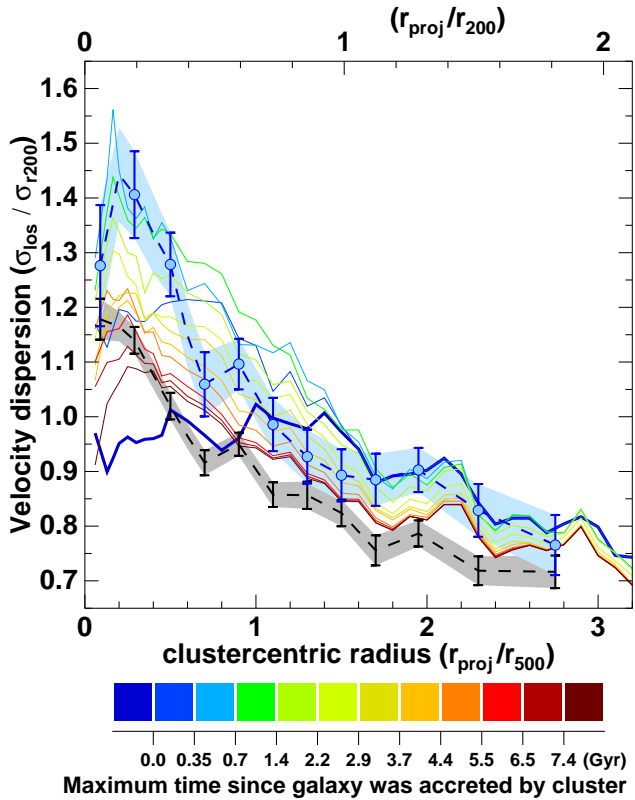}}
\caption{Predicted velocity dispersion profiles as a function of accretion epoch for model galaxies in our stacked sample of 75 clusters from the Millennium simulation at $z{=}0.21$. The thick blue curve shows the predicted LOS velocity dispersion profile of ``spectroscopic'' cluster members yet to pass within $r_{200}$ and be accreted into the cluster DM halo. The remaining solid curves show the impact on the velocity dispersion of progressively including those galaxies accreted into the clusters within the last $\Delta t$\,Gyr as indicated by the color scale. The black solid curve shows the velocity dispersion profile obtained by including all cluster members. Blue symbols with error bars show the {\em observed} velocity dispersion profile of the 24$\mu$m-detected cluster members from our stacked sample of 30 clusters. The black dashed curve shows the corresponding profile considering all $\mathcal{M}{>}2{\times}10^{10}\,{\rm M}_{\odot}$ cluster members.
}
\label{vel_profile}
\end{figure}

The velocity dispersion profile (VDP) of cluster galaxies, $\sigma_{los}(r)$, 
provides complementary constraints for the accretion epochs of galaxy sub-populations (see Fig.~\ref{zspace_stacked}d). 
Figure~\ref{vel_profile} compares the observed VDPs of 24$\mu$m-detected star-forming galaxies ({\em blue points}) with the predicted VDPs of model cluster galaxies, selected according to their accretion epoch. 

The VDP for star-forming cluster galaxies shows a high, narrow peak of $1.44\,\sigma_{\nu}$ at $r_{proj} \sim 0.3\,r_{500}$, before dropping to the innermost radial bin, and a steady decline outwards to ${\sim}0.8\,\sigma_{\nu}$ at large radii. 
This profile shape is best reproduced by model cluster populations combining infalling galaxies yet to be accreted and the most recent arrivals into the cluster. The progressive inclusion of these recently accreted galaxies causes the velocity dispersion within $r_{500}$ to rise rapidly, producing a characteristic sharp peak at $r_{proj}{\sim}0.2\,r_{500}$ which reaches a maximum height of ${\sim}1.56\,\sigma_{\nu}$ when $\Delta t = 0.7$\,Gyr, along with a corresponding sharp drop off to the cluster core. The best overall match to observations is produced for models with $\Delta t {\sim}0$.5--2.2\,Gyr ({\em light blue/green curves}),  
comparable with the time-scales required for infalling galaxies to approach the pericenter of their orbits through the cluster and achieve the high velocities required to produce the observed peak in $\sigma_{los}(r)$.
The observed profile is inconsistent with models with much longer delays between accretion into the cluster and quenching ($\Delta t {\ga}3.7$\,Gyr), and models in which star-formation is quenched instantaneously upon accretion (${\Delta t}{=}0$\,Gyr; {\em thick blue curve}), due to their predicted low, relatively flat LOS VDPs within $r_{500}$.

 The observed VDP considering all cluster members ({\em black dashed curve}) is not well matched by any model profile. While the predicted VDP considering all ``spectroscopic'' cluster galaxies peaks at ${\sim}0.2\,r_{500}$ and drops off sharply towards the cluster core ({\em dark red curve}), the observed VDP shows no corresponding dip in the cluster core. The $\sigma_{los}(r)$ instead declines steadily from its peak value 1.18$\,\sigma_{\nu}$ in the innermost radial bin, falling to values of ${\sim}0.75\,\sigma_{\nu}$ over the range 1.6--2.$8\,r_{500}$, significantly below the velocity dispersions predicted by simulations at these radii. 
Our observed VDP is qualitatively similar to that obtained by \citet{rines03} by stacking the member galaxies of eight $z{<}0.05$ X-ray luminous clusters: their $\sigma_{los}(r)$ also drops from 1.1$\sigma_{\nu}$ in the cluster core to $0.8\,\sigma_{\nu}$ by $r_{200}$, albeit with marginal evidence for a decline within $0.1\,r_{200}$. At larger radii (${\ga}2\,r_{500}$), their $\sigma_{los}(r)$ drops to values of ${\sim}0.5\,\sigma_{\nu}$, which is even lower than ours and hence poses further problems for the simulations.  

One possible explanation is that the mis-match is linked to the prediction of too many model galaxies in the cluster core ($r_{proj}\la 0.1\,r_{500}$; Fig.~\ref{log_profile}), which assuming that the excess population were all accreted early, could artificially increase the contribution from low-velocity virialized cluster members, reducing the $\sigma_{\nu}$ estimates for each cluster, and pushing the resultant model curves upwards. 

Irrespective of the difficulties in reproducing the observed VDP of all cluster members, the key finding that the velocity dispersion of star-forming galaxies is 10--35\% higher than that of the overall cluster population at all radii, along with the apparent sharp peak in the VDP at ${\sim}0.3\,r_{500}$, unambiguously identifies the star-forming cluster galaxy population as recent arrivals. 
Considering a \mbox{simple kinematical} treatment of infalling and virialized cluster galaxies in a cluster-scale gravitational potential well leads to $|T/V|{\approx}1$ for infalling galaxies and $|T/V|{\approx}1/2$ for the virialized population, where $T$ and $V$ are the kinematic and potential energies. Thus, the velocity dispersions of the two populations are naively related by $\sigma_{infall}{\approx}\sqrt{2}\,\sigma_{virial}$ \citep{colless}. 

From the first dynamical studies of cluster galaxies, the velocity dispersions of spiral galaxies have been found to be systematically higher than early types \citep{tammann, moss}. Based on much larger samples, the stacked velocity dispersions of blue/emission-line galaxies were found to be 20\% higher than the remaining inactive galaxies \citep{biviano97,aguerri}. \citet{biviano} showed that if early-type galaxies are assumed to have isotropic orbits within clusters, as supported by their Gaussian velocity distributions, the kinematic properties of late-type spirals are inconsistent with being isotropic at the $>99$\% level. Instead they indicate that spirals and emission-line galaxies follow radial orbits in clusters, pointing towards many of them being on their first cluster infall. 

\begin{figure}
\centerline{\includegraphics[width=84mm]{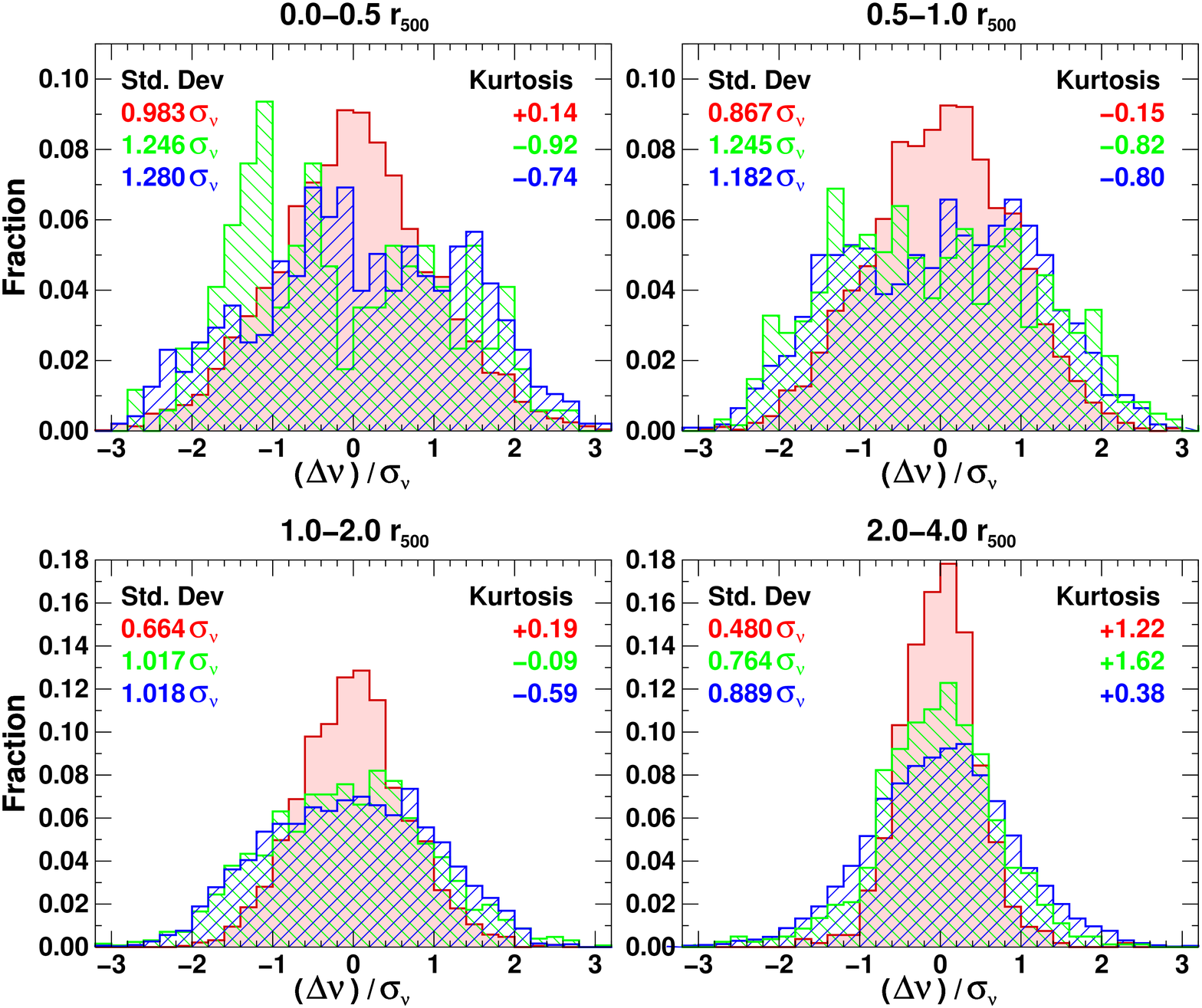}}
\caption{Stacked relative LOS velocity distributions of ``spectroscopic'' cluster galaxies for the 75 most massive clusters in the Millennium simulation at $z{=}0.21$ split into three dynamic sub-populations: (i) infalling ``spectroscopic'' cluster members yet to reach pericenter ({\em blue striped histogram}); (ii) back-splash galaxies which have passed through pericenter, are on their way back out of the cluster $\nu_{radial}{>}0$, and are yet to reach apocenter ({\em green striped histogram}); and (iii) the virialized cluster population which has passed through apocenter ({\em red solid histogram}). Each panel represents a different bin in cluster-centric radius as in Fig.~\ref{vel_dists}. The standard deviations (in units of $\sigma_{\nu}$) and kurtosis values of each distribution are indicated.} 
\label{vel_dists_model}
\end{figure}

Figure~\ref{vel_dists} showed that the LOS velocity distribution of star-forming cluster galaxies within $r_{500}$ to have a rather flat, top-hat profile, a high LOS velocity dispersion and a negative kurtosis ($\gamma_{2}{\sim}-0.81$) strongly inconsistent with the Gaussian distribution typical of a virialized cluster population. This flat-topped distribution is well reproduced by model cluster galaxy populations that are either infalling into the cluster for the first time, or back-splash galaxies which are currently rebounding out of the cluster and are yet to reach apocenter (Fig.~\ref{vel_dists_model}). The velocity distributions of these two dynamical sub-populations appear indistinguishable within $r_{500}$, both having velocity dispersions ${\sim}1.2\sigma_{\nu}$ and negative kurtosis values $\gamma_{2}{\sim}{-}0.8$. 
At 1--2$\,r_{500}$ the velocity distribution of star-forming galaxies becomes more rounded, albeit still with a negative kurtosis ($\gamma_{2}{\sim}-0.6$), which again is well reproduced by the model infalling galaxy population ($\gamma_{2}{=}-0.59$). At these radii, the back-splash population is expected to show a more Gaussian-like distribution with $\gamma_{2}{=}-0.09$, inconsistent with observations. However, there are only expected to be 40\% as many back-splash galaxies in this radial bin as infalling ones, and so we cannot rule out the possibility these star-forming galaxies represent a mixture of infalling and back-splash populations.

\subsection{Distribution of galaxies in the caustic diagram}

\begin{figure}
\centerline{\includegraphics[width=70mm]{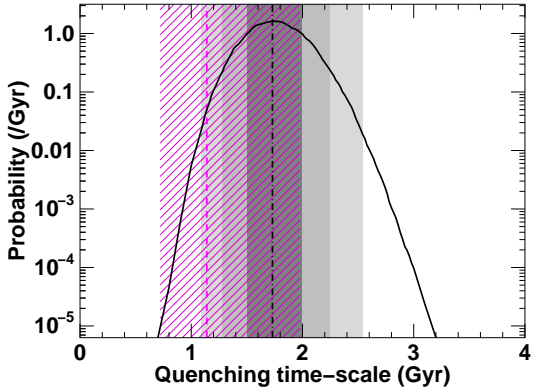}}
\caption{Likelihood function of the quenching time-scale, $t_{Q}$, based on fitting the distribution of star-forming galaxies in the caustic diagram, normalized so that $\int\mathcal{L}(t_{Q}) = 1$ with $t_{Q}$ in units of Gyr. The best-fit value of $t_{Q}$ is indicated by the vertical dot-dashed line, while the shaded regions indicate the 1, 2 and 3-$\sigma$ confidence limits in $t_{Q}$. The magenta dashed line and hashed region indicates the best-fit $t_{Q}$ value and 1-$\sigma$ confidence limits by \citet{haines13} based on the systematically low specific-SFRs of star-forming cluster galaxies within $r_{200}$.}
\label{likelihood}
\end{figure}
 
The information gained from the galaxy surface density profiles and VDPs can be combined by comparing the observed spatial distribution of star-forming galaxies in the stacked caustic diagram (Fig.~\ref{sf_zspace}) with those obtained from the Millennium simulation (Fig.~\ref{zspace_stacked}). 
Using the stacked sample of model galaxies from the 75 clusters extracted from the Millennium simulation, the distribution of model star-forming galaxies in the phase-space diagram $\Delta \nu_{los}/\sigma_{\nu}$ versus $r_{proj}/r_{500}$ was determined using the adaptive kernel estimator, for a range of quenching time-scales from 0--10\,Gyr. 

As in Section~\ref{sec:profiles} model star-forming cluster galaxies are given initial SFRs taken at random from our observed sample of coeval field star-forming galaxies (${\rm SFR}{>}2\,{\rm M}_{\odot}\,{\rm yr}^{-1}$; $0.15{<}z{<}0.30$). These are then set to decline exponentially on a quenching time $t_{Q}$ once they pass within $r_{200}$ of the cluster, until their SFRs fall below our nominal limit of 2.0\,M$_{\odot}\,{\rm yr}^{-1}$. 
The redshift at which each model galaxy was accreted into the cluster is known, from which the number of quenching time-scales $t_{Q}$ passed between the epoch of accretion and the epoch of observation ($z{=}0.21$) can be determined, and hence its final SFR and whether it would still be classified as star forming. Those galaxies yet to be accreted are assumed to still be identified as star forming. 
For each quenching time-scale, $t_{Q}$, the spatial distribution of star-forming model galaxies in phase-space, $\rho(r_{proj}/r_{500},\Delta \nu_{los}/\sigma_{\nu})$, is determined using the adaptive Gaussian kernel method. Each model galaxy $i$ is represented by a 2D Gaussian kernel of width $\sigma_{0}(\rho_{i}/\overline{\rho})^{-1/2}$, where $\overline{\rho}$ is the geometric mean of the local densities $\rho_{i}$ of the model galaxies in phase-space, and $\sigma_{0}$ is the initial kernel width 0.2. To remove the effects of the discontinuity at $r_{proj}{=}0$, the phase-space distribution is mirrored about both velocity and radial axes. 
The distribution is normalized to unity when summed over the region $0.0{<}(r_{proj}/r_{500}){<}3.2$, so that it can be considered a probability distribution function, P($r_{proj}/r_{500}$, $\Delta \nu_{los}/\sigma_{\nu}$).

The probability distributions of model star-forming galaxies for each value of $t_{Q}$ are then compared with the observed distribution of star-forming galaxies in the stacked phase-space diagram (Fig.~\ref{sf_zspace}).  
The best-fitting model cluster population is identified using a maximum-likelihood analysis, determining the value of $t_{Q}$ for which the likelihood $\mathcal{L}{=}\prod_{i=1}^{N} P(r_{i,proj}/r_{500},\Delta\nu_{i}/\sigma_{\nu}|t_{Q})$ is maximized, taking into account corrections for spectroscopic incompleteness and radial variation in coverage by our 24$\mu$m images as before. 

Figure~\ref{likelihood} displays the resulting likelihood function $\mathcal{L}(t_{Q})$ as well as the 1, 2 and 3-$\sigma$ confidence limits in $t_{Q}$.
The closest match to the observed distribution of star-forming galaxies is obtained for a value of $t_{Q}{=}1.73{\pm}0.25$\,Gyr. Quenching time-scales below 1\,Gyr and above 3\,Gyr are both excluded at ${>}3\sigma$ level, primarily due to the radial distribution of star-forming galaxies observed in our clusters (Fig.~\ref{profile_exp}). 

\begin{figure}
\centerline{\includegraphics[width=84mm]{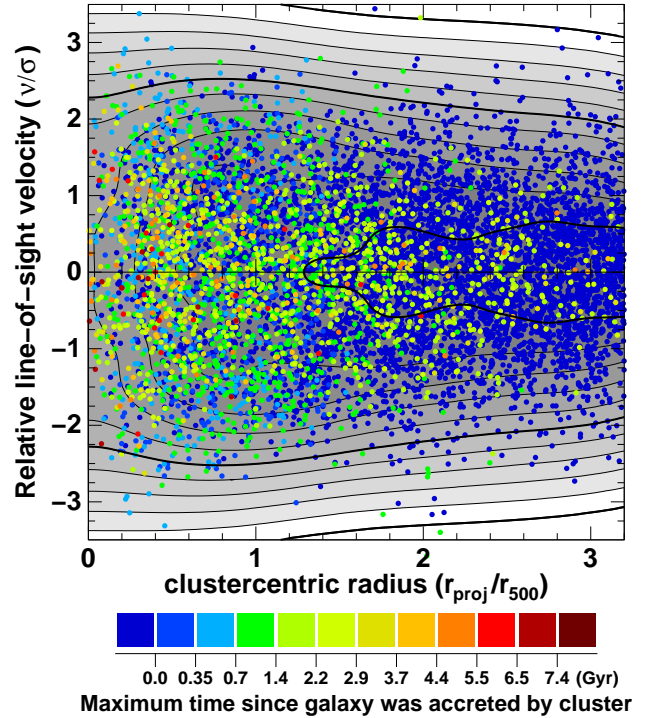}}
\caption{Stacked phase-space diagram, $\Delta \nu_{z}/\sigma_{\nu}$ versus $r_{proj}/r_{500}$, for model star-forming galaxies around the 75 most massive clusters in the Millennium simulation at $z{=}0.21$ for the best-fit quenching model, in which star formation declines exponentially on a time-scale $t_{Q}{=}1.73$\,Gyr in galaxies from the moment they pass within $r_{200}$ of the cluster for the first time. Each point marks a $\mathcal{M}{>}2{\times}10^{10}\,{\rm M}_{\odot}$ star-forming galaxy with ${\rm SFR}{>}2.0{\rm M}_{\odot}{\rm yr}^{-1}$, colored according to time elapsed since it was accreted into the cluster (mid-blue if it is yet to be accreted). The grayscale contours indicate the resulting probability distribution function of the model star-forming galaxies in phase-space, P($r_{proj}/r_{500},\Delta \nu_{z}/\sigma_{\nu}$). Each contour indicates a factor 0.2\,dex change in phase-space density of model star-forming galaxies. The thick contours indicate a factor 10 change in phase-space density.} 
\label{quenching_caustics}
\end{figure}

Figure~\ref{quenching_caustics} shows the corresponding stacked caustic diagram of model star-forming cluster galaxies for $t_{Q}{=}1$.73\,Gyr. Each cluster galaxy is color coded according to its accretion epoch, and the overall phase-space density distribution is shown by the grayscale contours. 
The model star-forming galaxies are most frequently found at large cluster-centric radii ($r_{proj}{\ga}1.5\,r_{500}$) and low velocity offsets ($|\Delta \nu_{los}|{\la}1.1\sigma_{\nu}$). The phase-space density of star-forming galaxies drops by a factor ${\sim}3{\times}$ as $r_{proj}/r_{500}\to 0$, for $|\Delta \nu_{los}|{\la}1.4\,\sigma_{\nu}$, producing the vertical contours in the left-center of the plot. 

Averaging over the 75 simulated clusters, 42\% \mbox{of the model} star-forming galaxies with $r_{proj}{<}r_{200}$ and line-of-sight velocities identifying them as ``spectroscopic'' cluster members are physically located outside $r_{200}$ at the time of observation ($z{=}0.21$) and have not been inside this radius at any time ({\em mid-blue points}), 33\% have been accreted within the last Gyr ({\em light blue/green points}), 19\% were accreted 1--3\,Gyr ago, and just 6\% were accreted ${>}3$\,Gyr prior to observation. With the model SFRs declining exponentially upon accretion on a time-scale of 1.73\,Gyr, overall the model SFRs and specific-SFRs of the star-forming cluster galaxies ($r_{proj}{<}r_{200}$) are reduced by 25\% from their values prior to accretion, consistent with the systematic reduction of 28\% in their specific-SFRs found in \citet{haines13}.
 
\section{Discussion}
\label{sec:discuss}

We have shown via two independent methods that star formation in galaxies infalling into clusters is not extinguished immediately upon their arrival into the cluster, but requires a significant time of the order 1--3\,Gyr to be quenched. First the steadily increasing surface density of star-forming galaxies towards the cluster core (Fig.~\ref{profile_exp}), as opposed to the flat radial profile predicted for instantaneous quenching models, simply requires star-forming galaxies to survive for a certain period within the cluster to build up the over-density seen in the cluster core, with a best-fit exponential quenching time-scale $t_{Q}{=}2.2{\pm}0.4$\,Gyr. Second, the velocity dispersion profile of star-forming cluster galaxies is consistently 10--35\% higher than inactive cluster galaxies at all radii, rising up to a sharp peak of $1.44\,\sigma_{\nu}$ at $0.3\,r_{500}$, which is inconsistent with instantaneous quenching models, but similar to the predicted VDPs of models in which star-forming galaxies survive for 0.5--2.2\,Gyr after being accreted (Fig.~\ref{vel_dists_model}). Combining both radial and velocity information, we compared the spatial distribution of star-forming galaxies within the caustic diagram with predictions from the Millennium simulation, to obtain a best-fit value of $t_{Q}{=}1.73{\pm}0.25$\,Gyr. 

While these results robustly confirm that star-forming galaxies are able to continue forming stars for some significant period after being accreted into massive clusters, it is not possible simply based upon the kinematics or distribution of star-forming galaxies within clusters to distinguish between slow quenching models whereby star-formation declines exponentially (or linearly) over a long time-scale, or a ``delayed-then-rapid'' quenching model in which recent arrivals continue to form stars normally for a certain period $\Delta t$, before suddenly stopping \citep{wetzel12}, as demonstrated by the similarity of the model curves in Figures~\ref{profile} and~\ref{profile_exp}. 

\subsection{The slow quenching of star formation in cluster galaxies}

The key distinguishing feature of the ``slow quenching'' model is its impact on the distribution of SFRs or specific-SFRs among star-forming cluster galaxies, systematically lowering the mean specific-SFRs as a significant fraction of star-forming cluster galaxies are observed during this process of slow quenching, while the ``delayed-then-rapid'' quenching model leaves the specific-SFR distribution of star-forming cluster galaxies unchanged. Figure~\ref{kinematics} supports the slow quenching model by finding kinematic segregation between star-forming cluster galaxies with normal or enhanced star formation, and those with {\em reduced} star-formation, indicative of ongoing quenching. This suggests that the process of quenching occurs over a sufficiently long time-scale that the kinematics and cluster-centric radii of {\em quenching} star-forming galaxies to have evolved significantly. 
More definitively, in \citet{haines13} we found that the specific-SFRs of massive ($\mathcal{\ga}10^{10}{\rm M}_{\odot}$) star-forming cluster galaxies within $r_{200}$ to be systematically 28\% lower than their counterparts in the field at fixed stellar mass and redshift, a difference significant at the 8.7$\sigma$ level. The entire specific-SFR distribution was seen to be shifted to lower values, marking the unambiguous signature of star formation in most (and possibly all) star-forming galaxies being {\em slowly} quenched upon their arrival into massive clusters. Assuming a model in which the SFRs decline exponentially upon passing within $r_{200}$, we obtained a best-fit quenching time-scale of $1.17_{-0.45}^{+0.81}$\,Gyr (magenta dashed line and hashed region in Fig.~\ref{likelihood}), consistent with the time-scales obtained here. 

As well as skewing the specific-SFR distribution of cluster galaxies on or near the star-forming main sequence, the slow quenching model is expected to result in significant numbers of galaxies with specific-SFRs of ${\sim}10^{-11}{\rm yr}^{-1}$ (or SFRs of 0.1--1\,M$_{\odot}\,{\rm yr}^{-1}$), well below the sensitivity of our {\em Spitzer} data, and filling in the ``green valley'' gap between star-forming and passive galaxies. \citet{wetzel13} found no evidence of this large transition population, leading them to prefer a delayed-then-rapid quenching model. One possible way of reconciling these two results would be a slow-then-rapid quenching model, whereby star formation is slowly quenched for the first 1--3\,Gyr, to explain the results of \citet{haines13}, followed-by a second short phase in which the residual star-formation is rapidly terminated, in order to retain the observed bimodal specific-SFR distribution of satellite/cluster galaxies \citep{haines11b,wetzel13}.

\citet{taranu} studied the bulge and disk colors of giant galaxies in $z {\le} 0.1$ clusters, finding shallow, gradual radial trends in disk colors that could be reproduced by slow quenching models similar to our own, but requiring slightly longer time-scales of  $t_{Q}{=}$3--3.5\,Gyr. They ruled out short (${\la}1$\,Gyr) quenching time-scales and ``delayed-then-rapid'' quenching models as both produced much larger and sharper radial changes in the median disk colors than observed. \mbox{A number} of other studies have also argued for relatively long time-scales ($\sim$1--4\,Gyr) for the quenching of star formation in recently accreted cluster galaxies \citep{balogh00,moran,finn08,vonderlinden,delucia,wetzel13}.  

\citet{muzzin} found a population of post-starburst galaxies in nine $z{\sim}1$ clusters, whose stacked spectrum could be fit by the rapid quenching ($t_{Q}{=}0.4_{-0.4}^{-0.3}$\,Gyr) of typical star-forming galaxies, and which traced a coherent ``ring'' at 0.25--0.50\,$r_{200}$ in the stacked caustic diagram that could be reproduced by recently accreted galaxies quenched 0.1--0.5\,Gyr after passing within 0.5\,$r_{200}$.
This much more rapid quenching in high redshift clusters could be due to the shorter gas consumption time-scales of galaxies at $z{\sim}1$ \citep{carilli} or ${\sim}1.7{\times}$ shorter cluster crossing time-scales. 

Slow quenching on ${\sim}2$\,Gyr time-scales matches predictions of starvation models, in which infalling galaxies are stripped of their diffuse gaseous halos as they pass through the ICM, preventing further gas accretion onto the galaxies from the surrounding inter-galactic medium \citep{larson, bekki, mccarthy}. The galaxy then slowly uses up its existing molecular and H{\sc i} gas reservoir over a period of 2--3\,Gyr, based on the gas consumption time-scales observed for nearby spiral galaxies \citep{bigiel,boselli14}. This process may be effective well beyond the virial radius, with the extended gaseous halo of clusters remaining hot (${\ga}10^{6}$K) and sufficiently dense to strip the hot gas atmospheres of infalling galaxies out to ${\sim}5\,r_{200}$ \citep{bahe,gabor}. 

Moreover, the growth of galactic DM halos is suppressed by tidal effects due to the presence of nearby cluster-mass halos \citep{hahn}, which results in the peak mass of galactic halos infalling into clusters occurring at ${\sim}1.8\,r_{vir}$ ($3.5\,r_{500}$) on average \citep{behroozi}. Given the tight correspondance between DM mass accretion and gas accretion onto galactic halos, including a significant fraction of gas accreted onto the galaxy via cold, dense filamentary streams at all redshifts \citep{vandevoort}, the radius of peak halo mass marks the end of continual gas replenishment of the ISM, signalling the beginning of the end for star formation in the host galaxy \citep{sanchez}. 

As these galaxies continue their journey towards the cluster core, the ICM they encounter becomes increasingly dense and the resultant ram pressures ($P_{ram}{\propto}\rho_{ICM}\nu^{2}$) become strong enough to progressively strip their gas disks from the outside in \citep{bruggen}, producing truncated H{\sc i}, H$_{2}$ gas and H$\alpha$ disks \citep[e.g.][]{koopmann,boselli06,boselli14}, and outer regions showing recently quenched stellar populations \citep{crowl}. Hydrodynamical simulations predict that the moderate ram pressures acting on infalling massive spirals as they pass within ${\sim}r_{500}-r_{200}$ are sufficient to strip half their gas contents on 500--1000\,Myr time-scales \citep{roediger}, while observations confirm that ram pressure stripping is effective at removing gas and quenching star formation as far out as $r_{500}$ \citep{chung07,merluzzi}. The effective time-scale for quenching via ram-pressure stripping then becomes the ${\sim}1$\,Gyr required for galaxies to travel from the cluster outskirts to the pericenter where the peak in ram pressure occurs \citep{roediger}. The complete absence of H{\sc i}-normal spirals within the $r_{500}$ radius of Virgo and Coma clusters \citep{boselli}, and the ${\sim}0$.5--0.8\,dex H{\sc i} deficiencies of Virgo spirals at fixed stellar mass, ${\rm NUV}-r$ color and stellar mass surface density, cannot be reproduced by predictions from chemospectrophotometric models involving starvation alone and can only be explained by ram-pressure stripping actively removing the gas \citep{cortese}. 

In typical spirals, the dust-to-gas ratio and internal extinction ($A_{FUV}$) decline steadily with radius \citep{munoz}, meaning that for ram pressure stripping events where gas is removed from the outside in, 
the more extended unobscured star formation component should be preferentially quenched prior to the more concentrated and bound obscured SF. This is supported by the finding of a significant cluster population of 24$\mu$m-detected spirals with reduced SFRs, indicative of ongoing slow quenching, reddened by dust onto the optical red sequence \citep{wolf}. This appears inconsistent with our finding that the quenching time-scale for UV-selected star-forming cluster galaxies is ${\sim}1$\,Gyr longer than that for 24$\mu$m-selected galaxies. We suggest that the longer $t_{Q}$ for the NUV-selected star-forming galaxies is due to the 1--1.5\,Gyr time-scale required for recently quenched galaxies to migrate from the blue cloud to red sequence in the $(NUV{-}r)$ versus $M_{r}$ C-M diagram \citep{kaviraj}, while the 24$\mu$m emission mostly originates from H{\sc ii} regions with ongoing star formation \citep[e.g.][]{calzetti}. 

\subsection{Pre-processing}
\label{sec:preprocessing}

In the SF--radius trends of Figs.~\ref{sf_radius}--\ref{UV_sf_radius}, the fraction of star-forming cluster galaxies rises steadily with radius, but crucially 
the $f_{SF}(r)$ remain stubbornly 20--30\% below that seen in coeval field populations even out at $5\,r_{500}$, well beyond the maximal distances back-splash galaxies are expected to rebound to. 
As a result, these trends cannot be reproduced by models in which star formation is quenched in infalling galaxies via processes that are only initiated when they are accreted into the clusters (Fig.~\ref{sf_radius_models}).
Our SF--radius trends are very similar to those of \citet{chung}, who analysed 69 $z{<}0.1$ clusters covered by both SDSS spectroscopy and WISE 22$\mu$m photometry, finding that the fraction of star-forming galaxies with $L_{IR}{>}4.7{\times}10^{10}L_{\odot}$ increases steadily with cluster-centric radius, but remains well below the field value even at ${\sim}3\,r_{200}$. \citet{vonderlinden} found that suppression of star-formation in cluster galaxies could be traced out to ${\sim}4r_{200}$. \citet{wetzel12} found their $f_{SF}(r)$ suppressed with respect to field values as far out as 10\,$r_{200}$ around cluster-mass halos (M$_{200}{>}10^{14}{\rm M}_{\odot}$), but interestingly saw this suppression entirely limited to satellite galaxies, with no evidence of suppression among centrals found beyond 1--2\,$r_{200}$.

The best way to reproduce the observed SF--radius trends appears to require a certain fraction of infalling galaxies to arrive onto the clusters having already been quenched (Fig.~\ref{sf_lower}). One commonly identified mechanism by which galaxies may be transformed at large distances from the cluster center is through ``pre-processing'' in infalling galaxy groups \citep[e.g.][]{kodama,fujita,berrier,mcgee09,dressler13}. 
Star formation is suppressed in group galaxies, with $f_{SF}$ values intermediate between those of field and cluster galaxy populations \citep{wilman05,bai,mcgee11,rasmussen} up to at least $z{\sim}1$ \citep{balogh11,ziparo}. The fraction of star-forming galaxies within groups declines steadily with increasing group mass (at fixed stellar mass and group-centric distance) and proximity to the group center \citep{weinmann,wetzel12,woo}. The fractions and specific-SFRs of star-forming galaxies within groups both show {\em accelerated} declines since $z{\sim}1$ with respect to the coeval field population \citep{popesso} and also ``filament-like'' environments with comparably high galaxy densities but no X-ray emission \citep{ziparo}, indicative of ongoing slow quenching within groups  \citep{balogh11}, comparable to that seen in our clusters \citep{haines13}. Galaxies in groups with masses ${\ga}10^{13}{\rm M}_{\odot}$ are H{\sc i}-deficient at fixed stellar mass and $NUV-r$ color, suggesting that ram-pressure stripping can remove atomic gas from ${\sim}L^{*}$ spirals in such groups \citep{catinella}. 

Galaxy groups are ubiquitous, and host ${\sim}5$0\% of galaxies in the local Universe \citep{mcgee09}.
The impact of environmental quenching in groups will be manifest also in a reduced global fraction of star-forming galaxies in our coeval field sample.  
Hence, in order to reproduce the observation that the fraction of star-forming galaxies in the infall regions of clusters (2--3\,$r_{200}$) remains significantly lower than that seen in coeval field galaxies, a specific form of ``pre-processing'' is required, in which the galaxies infalling into clusters are {\em more likely} to be existing members of groups and hence {\em more likely} to be ``pre-processed'' than the cosmic average. That is we require the mass function of DM halos hosting galaxies in the surroundings of clusters (not including the cluster itself) to be top heavy and biased towards group-scale masses with respect to the cosmic average, as expected in such over-dense regions of the universe \citep{faltenbach}.

\begin{figure}
\centerline{\includegraphics[width=84mm]{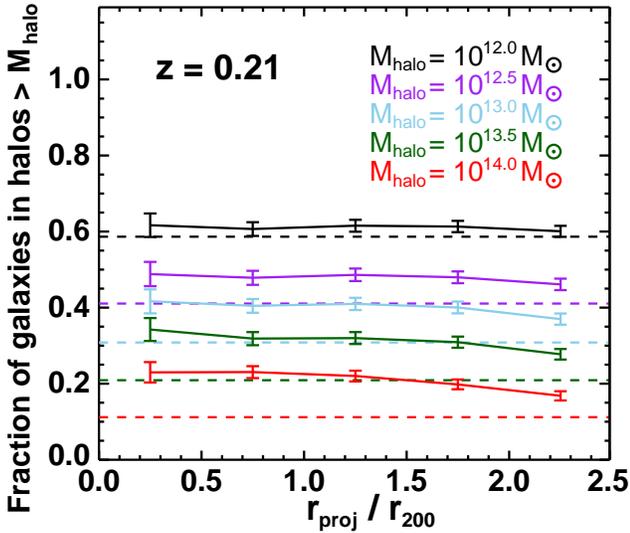}}
\caption{History bias for galaxies falling into massive clusters. The solid curves show the fraction of model galaxies surrounding massive clusters (M$_{cl}{>}10^{14.2}\,{\rm M}_{\odot}$) at $z{=}0.21$ which already reside in halos over a given mass as a function of projected cluster-centric radius. Only those galaxies yet to pass within $r_{200}$ and be accreted into the cluster are included. The dashed curves show the corresponding cosmic fractions for field galaxies over the whole Millennium simulation.} 
\label{preprocessing}
\end{figure}

To examine this issue, we construct the mass function of the DM halos hosting galaxies at different cluster-centric radii and compare it with that of coeval field galaxies. We select galaxies within 4000\,km\,s$^{-1}$ of the mean redshifts of rich clusters (M$_{\rm halo}{>}10^{14.2}{\rm M}_{\odot}$) in the Millennium simulation at $z{=}0.21$ \citep[see][for details]{lu}, excluding those galaxies already accreted into the clusters. Figure~\ref{preprocessing} shows the resulting fractions of these model galaxies which reside in halos over a given mass as a function of projected cluster-centric radius ({\em solid curves}) for five values of minimum halo mass from $10^{12}{\rm M}_{\odot}$ to $10^{14}{\rm M}_{\odot}$. The dashed lines show the corresponding fractions for coeval field galaxies over the whole Millennium simulation. This confirms that galaxies infalling into clusters are indeed more likely to already reside within DM halos of a given mass than coeval field galaxies, for each halo mass range. While the effect is marginal for  $10^{12}{\rm M}_{\odot}$ mass halos, galaxies infalling into massive clusters are ${\sim}3$5\% more likely to already be residing in group-mass halos with $M_{\rm halo}{>}10^{13}{\rm M}_{\odot}$ and twice as likely to already be residing in $M_{\rm halo}{>}10^{14}{\rm M}_{\odot}$ halos than their counterparts in the field.   

We find numerous potential sites where infalling galaxies can be pre-processed before being accreted into the LoCuSS clusters \citep[e.g.][]{pereira10}. 
Across the 23 clusters with {\em XMM} imaging, a total of 30 X-ray groups (with extended X-ray emission detected at ${>}4\sigma$ levels) have been identified with redshifts placing them inside the cluster caustics, indicating that they are most likely infalling into the primary cluster (Haines et al. 2015, in preparation). Only six further ``isolated'' X-ray groups were detected (at ${>}4\sigma$) in the same {\em XMM} images over the rest of the 0.15--0.30 redshift range.

The observed ratio of ${\sim}$5 X-ray groups associated with the clusters for every field X-ray group in the remainder of the $0.15{<}z{<}0.30$ volume covered by {\em XMM}, is double (quadruple) the ratio obtained for cluster and field $M_{K}{<}M^{*}_{K}{+}1.5$ (24$\mu$m-detected) galaxies in the same two volumes. This suggests that infalling galaxies are 2--4${\times}$ more likely to be members of X-ray groups than in typical field regions, assuming that both group samples have similar numbers of members per group. 

A second factor in the shortfall in $f_{SF}$ at large radii could arise from a bias in the mass assembly history of the infalling galaxies and groups themselves. \citet{maulbetsch} find that galactic ${\sim}10^{12}{\rm M}_{\odot}$ halos in high-density regions, such as the infall regions of massive clusters, form earlier, have more active merger histories, and have much lower mass accretion rates \citep[and commensurate gas accretion rates;][]{vandevoort} at late epochs ($z{\la}0.5$) than those which form in low-density field environments. Such halos in high-density regions are ${\sim}4{\times}$ likelier to not be accreting any mass (or even losing it) at late epochs than those in low-density regions. 

\section{Summary}
\label{sec:summary}

We present an analysis of the radial distribution and kinematics of star-forming galaxies in 30 massive clusters at $0.15{<}z{<}0.30$, combining wide-field {\em Spitzer} 24$\mu$m and {\em GALEX} NUV photometry with highly-complete spectroscopy of cluster members. 
To gain insights into how the observed trends relate to the continual accretion of star-forming spirals onto massive clusters and subsequent quenching of star formation, we follow the infall and orbits of galaxies in the vicinity of the 75 most massive clusters in the Millennium cosmological simulation, obtaining a series of predicted model trends that should have general applicability for understanding galaxy evolution in cluster environments. Our main results are summarized below: 

\begin{enumerate}
\item The surface density of star-forming galaxies declines steadily with radius, falling ${\sim}15{\times}$ from the cluster core to $2\,r_{200}$. This simple observation requires star formation to survive within recently accreted spirals for 2--3\,Gyr to build up the apparent over-density of star-forming galaxies within clusters.

\item The velocity dispersion profile of the star-forming cluster galaxy population shows a sharp peak of 1.44\,$\sigma_{\nu}$ at 0.3\,$r_{500}$, and is 10--35\% higher than that of the inactive cluster members at all cluster-centric radii, while their velocity distribution shows a flat, top-hat profile within $r_{500}$. All of these results are consistent with star-forming cluster galaxies being an infalling population, but one that must also survive ${\sim}0$.5--2\,Gyr beyond passing within $r_{200}$ to achieve the high observed velocities.

\item The distribution of star-forming galaxies in the stacked caustic diagram are best-fit by models in which their SFRs decline exponentially on quenching time-scales $t_{Q}{=}1.73{\pm}0.25$\,Gyr upon accretion into the cluster. The above results, and the observed kinematic segregation of star-forming galaxies according to their specific-SFRs, support the conclusion from \citet{haines13} that star formation in most (and possibly all) high-mass star-forming galaxies is {\em slowly} quenched on accretion into rich clusters on 0.7--2.0\,Gyr time-scales

\item The fraction ($f_{SF}$) of star-forming cluster galaxies rises steadily with cluster-centric radius, increasing five-fold by $2\,r_{200}$, but remains well below field values even at $3\,r_{200}$. Pre-processing in infalling galaxy groups appears the most likely explanation for this  suppression of star-formation at large distances from the cluster.
\end{enumerate}

\section*{Acknowledgements}

CPH was funded by CONICYT Anillo project ACT-1122. 
GPS acknowledges support from the Royal Society. FZ and GPS acknowledge support from 
the Science and Technology Facilities Council. 
We acknowledge NASA funding for this project under the Spitzer program GO:40872.  
This work was supported in part by the National Science Foundation under Grant No. AST-1211349. 
The Millennium simulation databases used in this paper and the web application providing online access to them were constructed as part of the activities of the German Astrophysical Virtual Observatory.

\label{lastpage}
\end{document}